\documentclass[prb,twocolumn,showpacs,preprintnumbers,amsmath,amssymb,floatfix,superscriptaddress]{revtex4}

\usepackage[]{graphicx}         	% Include figure files
\usepackage{bm}             		% bold math
\usepackage{tabularx}
\usepackage{times}				% Times font

\begin{document}

%%%%%%%%%%%%%%%%%%%%%%%%%%%%%%%%%%%%%%%%
%
%       Header information
%
%%%%%%%%%%%%%%%%%%%%%%%%%%%%%%%%%%%%%%%%
%\title{Lateral size dependence of spin wave modes in magnetic tunnel junction nanopillars}
\title{Quantized spin wave modes in magnetic tunnel junction nanopillars}

\author{A. Helmer}
\email{annerose.helmer@u-psud.fr}
\affiliation{Institut d'Electronique Fondamentale, UMR CNRS 8622, 91405 Orsay, France}
\affiliation{Universit{\'e} Paris-Sud, 91405 Orsay, France}
\author{S. Cornelissen}
\affiliation{IMEC, FNS, Kapeldreef 75, 3001 Leuven, Belgium}
\affiliation{ESAT, KU Leuven, Leuven, Belgium}
\author{T. Devolder}
\affiliation{Institut d'Electronique Fondamentale, UMR CNRS 8622, 91405 Orsay, France}
\affiliation{Universit{\'e} Paris-Sud, 91405 Orsay, France}
\author{J.-V. Kim}
\affiliation{Institut d'Electronique Fondamentale, UMR CNRS 8622, 91405 Orsay, France}
\affiliation{Universit{\'e} Paris-Sud, 91405 Orsay, France}
\author{W. van Roy}
\affiliation{IMEC, FNS, Kapeldreef 75, 3001 Leuven, Belgium}
\author{L. Lagae}
\affiliation{IMEC, FNS, Kapeldreef 75, 3001 Leuven, Belgium}
\affiliation{Natuurkunde en Sterrenkunde, KU Leuven, Leuven, Belgium}
\author{C. Chappert}
\affiliation{Institut d'Electronique Fondamentale, UMR CNRS 8622, 91405 Orsay, France}
\affiliation{Universit{\'e} Paris-Sud, 91405 Orsay, France}

\date{\today}

%%%%%%%%%%%%%%%%%%%%%%%%%%%%%%%%%%%%%%%%
%
%       Abstract
%
%%%%%%%%%%%%%%%%%%%%%%%%%%%%%%%%%%%%%%%%
\begin{abstract}
We present an experimental and theoretical study of the magnetic field dependence of the mode frequency of thermally excited spin waves in rectangular shaped nanopillars of lateral sizes $60\times100$, $75\times150$, and $105\times190~\textrm{nm}^2$, patterned from MgO-based magnetic tunnel junctions. The spin wave frequencies were measured using spectrally resolved electrical noise measurements. In all spectra, several independent quantized spin wave modes have been observed and could be identified as eigenexcitations of the free layer and of the synthetic antiferromagnet of the junction. Using a theoretical approach based on the diagonalization of the dynamical matrix of a system of three coupled, spatially confined magnetic layers, we have modeled the spectra for the smallest pillar and have extracted its material parameters. The magnetization and exchange stiffness constant of the CoFeB  free layer are thereby found to be substantially reduced compared to the corresponding thin film values. Moreover, we could infer that the pinning of the magnetization at the lateral boundaries must be weak. Finally, the interlayer dipolar coupling between the free layer and the  synthetic antiferromagnet causes mode anticrossings with gap openings up to 2 GHz. 
At low fields and in the larger pillars, there is clear evidence for strong non-uniformities of the layer magnetizations. In particular, at zero field the lowest mode is not the fundamental mode, but a mode most likely localized near the layer edges.
\end{abstract}
%%%     PACS, the Physics and Astronomy Classification Scheme.

\maketitle

%%%%%%%%%%%%%%%%%%%%%%%%%%%%%%%%%%%%%%%%
%
%                Paper
%
%%%%%%%%%%%%%%%%%%%%%%%%%%%%%%%%%%%%%%%%

\section{Introduction}

In the last few years, magnetic tunnel junction (MTJ) nanopillars have received tremendous attention due to their promising potential for applications in spin-transfer-switched Magnetic Random Access Memory or as spin-torque oscillators for microwave generation.\cite{Katine:JMMM:2008, Chappert:Nature:2007, Deac:Nature:2008} With GHz frequencies the operation speed of these devices happens to lie in the same frequency range as the dynamic eigenexcitations of the underlying nanoelements (thermally excited spin waves), which may therefore manifest themselves as unwanted noise sources. However, as eigenexcitations, thermal spin waves also constitute an excellent probe for the intrinsic magnetic properties of the nanopillars. The experimental detection of spin waves in MTJ nanopillar devices and the understanding of their nature is therefore of great interest both for fundamental and technological reasons.\\

Spin waves in confined structures have been studied extensively in single-layer dots with thicknesses between 40 and 15~nm and typical lateral dimensions from $3$~$\mu$m down to 200~nm.\cite{Bayer:book:2006,  Bailleul:PRB:2006, Gubbiotti:JMMM:2007, Gubbiotti:PRB:2005, Montoncello:PRB:2007} In these systems, two types of spin wave modes have been identified: quantized volume modes located around the center of the element where the internal field is basically homogeneous, and spin wave well or end modes localized near the element edges in the inhomogeneity region of the internal field. The above elements are characterized by their thickness being significantly larger than the exchange length of the layer material (typically 5~nm). In structures with this property the dominating interaction is the magneto-static dipolar interaction,\cite{Guslienko:PRB:2005} which causes the inhomogeneity of the internal field, thus determining the character and spatial profile of the modes.\cite{Bayer:book:2006} \\

The eigenexcitations of a multi-layer dot differ in general significantly from those of an ensemble of isolated magnetic dots due to the interlayer interactions between the magnetic layers in the stack: mutual dipolar coupling and - for sufficiently thin metallic spacer layers - interlayer exchange coupling.\cite{Vavassori:JPDApplPhys:2008}\\
Eigenexcitations of nanopillar structures have been the subject of very few studies so far. Thermal spin waves have been investigated systematically only in pseudo-spin-valves\cite{Gubbiotti:PRB:2006, Gubbiotti:MMMINT:2007, Vavassori:JAP:2008} of circular and elliptical shape (smallest dimension 200~nm) consisting of two magnetic layers of 10~nm thickness separated by a 10~nm thick spacer layer, i.e. again layer thicknesses were much larger than the exchange length. Consequently, the  profiles of the modes in each of the two pillar layers showed great resemblance\cite{Gubbiotti:PRB:2005,Gubbiotti:MMMINT:2007} with the mode profiles in the corresponding isolated dots. In a symmetric spin-valve stack,\cite{Gubbiotti:PRB:2006} the main impact of the mutual dipolar coupling between the layers  was found to be a fixed phase relation between the modes in the two layers for high applied field, and hybridization effects at low field.\\

Common MTJ nanopillars differ qualitatively from the pseudo-spin-valves in three fundamental points: Firstly, with a free layer and a synthetic antiferromagnet (SAF) they consist of three magnetic layers; secondly, with $2-4$~nm the layer thicknesses are now smaller than the exchange length,\cite{Bilzer:JAP:2006} such that, for sufficiently small lateral dimensions, the spin dynamics in each layer is dominated by the exchange interaction; thirdly, the interlayer interaction of the three layers is highly asymmetric: the two SAF layers are strongly coupled by interlayer exchange and - more weakly - mutual dipolar coupling, one of them (the pinned layer) being additionally subject to the strong exchange bias field; the free layer interacts with the SAF via the comparatively weak mutual dipolar coupling only. Consequently, the eigenexcitations of an MTJ nanopillar are expected to be much more complex than those of the pseudo-spin-valves. \\

In this paper, we investigate the magnetic field dependence of the mode frequency of thermally excited spin waves in rectangular shaped MgO-based MTJ nanopillars of different lateral sizes. In section II, we will describe the basic magnetic properties of the devices and the experimental techniques used to acquire the spin wave spectra. The features of the measured spectra in dependence of the pillar size and the direction of the applied field are described in the following section III. In section IV, we will point out short-comes of the macrospin model when applied to our samples, as a consequence of which we will introduce in section V a model of quantized spin wave modes in nanopillars consisting of three magnetic layers. In section VI, we will use this model to extract the material parameters of the pillar, which will finally be discussed in section VII along with the limitations of our model.

\section{Samples and experimental techniques}
\label{sec:exp}
\subsection{Samples and basic device properties}
The fabrication and basic properties of our samples are described in Ref.~\onlinecite{Cornelissen:JAP:2009}: they are rectangular shaped nanopillars, all patterned from the same MTJ stack of composition $\textrm{Co}_{60}\textrm{Fe}_{20}\textrm{B}_{20}$ (3~nm, free layer)/ $\textrm{Mg} (1.3)$[nat.~ox.]/ $\textrm{Co}_{60}\textrm{Fe}_{20}\textrm{B}_{20}$ (2, reference layer)/ $\textrm{Ru} (0.8)/ \textrm{Co}_{70}\textrm{Fe}_{30}$ (2, pinned layer)/$\textrm{PtMn} (20)$, deposited by Singulus Technologies AG. The three layers following the MgO tunnel barrier compose the synthetic antiferromagnet (SAF). The pillars were designed in three lateral sizes: $60 \times 100$, $75 \times 150$, and $105 \times 190~\textrm{nm}^2$, which will be referred to as small (S), medium (M), and large (L) size, respectively. Note that unlike in Ref.~\onlinecite{Cornelissen:JAP:2009} the given dimensions are not the nominal values, but mean values measured on the exposed e-beam resist with a device-to-device deviation of $\pm10$~nm. In order to obtain electrically contactable devices the nanopillars were inserted in series between coplanar leads, following design rules ensuring high bandwidth.\cite{Devolder:JAP:2008}\\

The devices have a resistance area product of  $16~\Omega\,\mu\textrm{m}^2$ and typically $80\%$  tunnel magneto-resistance ratio. Their hysteretic properties are consistent with the uniaxial anisotropy expected from the rectangular pillar shape where the long edge of the rectangle, oriented along the exchange pinning direction of the PtMn antiferromagnet, is the easy axis (EA), and the short edge the hard axis (HA) of the magnetization. Panels (b) and (d) of Fig.~\ref{fig1} show as a reference EA and HA hysteresis loops of a nanopillar of size S calculated in macrospin approximation using as material parameters literature bulk values (see figure caption). In comparison, the experimental EA and HA loops for devices of pillar size S, M, and L are depicted in panels (b),~(d),~and~(f) of Figs.~\ref{fig2},\ref{fig3}, respectively.\\ 
At negative EA applied fields the pillars are in the parallel (P) state, at positive fields in the antiparallel (AP)  state; spin-flop (SF) transition of the SAF occurs typically at EA fields around $+170$~mT. Room temperature coercivity is $25-35$~mT for devices of size S and M, and $20-25$~mT for size L. From astroid measurements\cite{Cornelissen:JAP:2009} mean anisotropy fields of 37~mT, 46~mT, and 38~mT for pillar sizes S, M, and L, respectively, have been determined. The EA loops of all devices are off-centered towards negative fields, indicating non-negligible antiparallel coupling of the free layer magnetization and the SAF. With increasing pillar size this coupling is decreasing: while for size S the shift is $5-11$~mT, it is only $3-7$~mT for size M, and $1-5$~mT for size L.\\
%%%%%%%%%%%%%%%%
The bell shape of the HA hysteresis loops (Fig.~\ref{fig3}) is consistent with the antiparallel coupling observed on the EA: At zero HA applied field the devices are always in the AP state. With increasing (absolute value of the) field the resistance decreases continuously from the maximum resistance of the AP state down to almost parallel remanence, as the magnetizations of the free layer and finally the two SAF layers progressively tilt towards the applied field. Ascending and descending field branch of the HA loops are for most devices identical. In contrast, in the absence of coupling between free layer and SAF, the pillar relaxes with equal probability into P state or AP state when the HA field is switched off, resulting in two branches of the hysteresis loop (see Fig.~\ref{fig1}(d)). Finally, note that the sharp bends in the resistance curve at about $\pm60$~mT in the measured HA loop for pillar size L, and also present in the calculated HA loop, become more and more rounded for decreasing pillar size, i.e. for increasing antiparallel coupling.\\

%%%%%%%%%%%%%%%%
We have used the intrinsic symmetry of the HA loops to align the external field with the symmetry axes of the rectangle by choosing the field direction such that the loops showed highest possible symmetry. For some devices the loops were, though symmetric at high fields, noticeably asymmetric at low fields; in these cases we have cross-checked the alignment with the symmetry of the corresponding HA spectra. The misalignment of the field should therefore not exceed $2^{\circ}$.

\subsection{Set-up and experimental methods}

To obtain their spin wave spectra the devices were inserted into a high bandwidth circuit similar to that in Ref.~\onlinecite{Devolder:PRB:2005}, and their voltage noise power spectrum density (PSD) was measured for moderate dc bias currents as a function of the applied magnetic field. The noise spectrum at each field step was obtained by subtracting from the spectrum for non-zero bias current a zero-current reference spectrum in order to eliminate noise of non-magnetic origin. In  panels (a),~(c),~and~(c) of Figs.~\ref{fig2},\ref{fig3} examples of 2D density plots of the PSD versus the magnetic field are shown, where the dark regions correspond to maxima in the PSD and therefore to eigenexcitations of the magnetic system,\cite{Smith:JAP:2001} i.e. spin wave modes. The spectra are displayed in a contrast scaling with the logarithm of the noise in excess to the noise at zero current. As the difference in amplitude between the most intense and the weakest modes is even on a logarithmic scale still large, in all figures the gray-scale of the PSD has been modulated, and black dots have been superimposed to better evidence the weaker modes.\\
Bias currents used to measure the spectra were chosen as low as possible in order not to affect the mode frequencies, but still high enough to obtain sufficient signal-to-noise ratio. Devices of pillar size S were therefore mostly measured at $\pm 0.1$~mA, those of size M at $\pm 0.2$~mA, and those of size L at $\pm 0.3$~mA. The differences in amplitude for opposite current polarity were hardly noticeable on the logarithmic scale, and the maximum difference in the mode frequencies, observed for 0.3~mA, was 0.15 GHz. These observations are in agreement with previous works\cite{Cornelissen:EPL:2009} on similar samples, where a spin-torque threshold current of 1.6~mA for size-L devices has been determined. On the frequency scales considered in this paper the bias current dependence of the spectra can therefore be neglected. \\

Note that our measurement technique allows to detect the spin wave modes of individual pillars, in contrast to the Brillouin Light Scattering technique used in Refs.~\onlinecite{Bayer:book:2006, Gubbiotti:PRB:2006, Vavassori:JAP:2008}, or the frequency-domain coplanar waveguide technique of Ref.~\onlinecite{Bailleul:PRB:2006}, where the measured spectra were an average over a large number of devices. Moreover, we do not need optical or any other direct access to the magnetic layers of the pillar, but can measure them in their natural working environment, i.e. as part of a stack used in (actual or potential) functional devices, and subject to electrical currents. Finally, 
since we measure the voltage noise, which is in one-to-one correspondence with the magneto-resistance (MR) noise, we are equally sensitive to spin waves in the free layer (FL) and the reference layer (RL).

\section{Experimental results}
\label{sec:results}
%**************************************************************************************
\begin{figure}
\includegraphics[width=8.5cm]{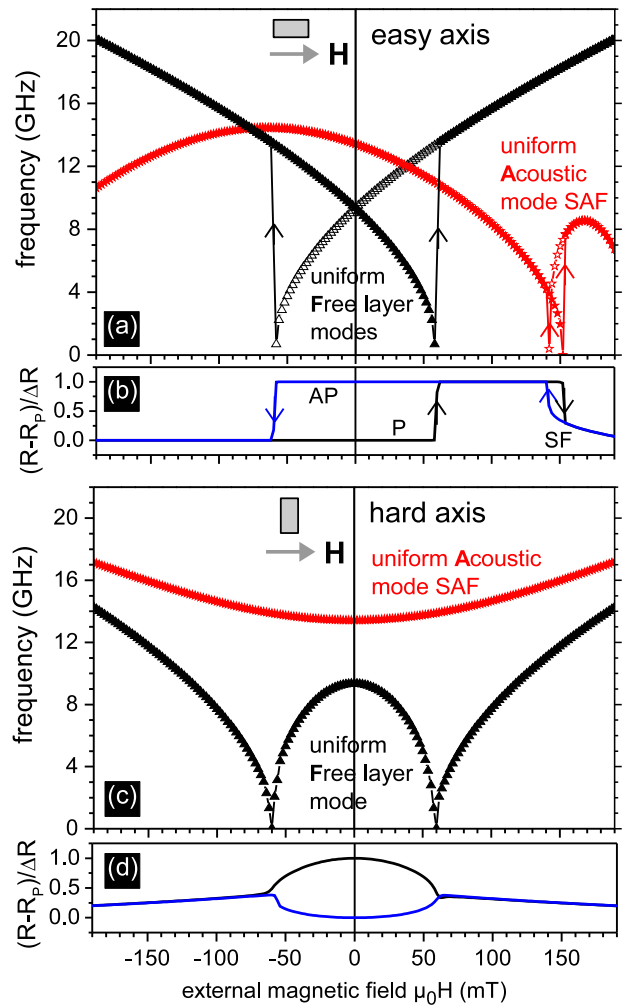}
\caption{\label{fig1}Macrospin description of a nanopillar of size S: frequencies of the uniform modes versus external field along (a) easy axis and (c) hard axis, calculated in macrospin approximation using as material parameters literature bulk or thin film values (see Appendix~\ref{sec:appendix-literature}): magnetization 2.2~T for CoFe, 1.9~T for annealed CoFeB, exchange bias $J^{eb}= 4.5 \times 10^{-4}\textrm{J}/\textrm{m}^2$, interlayer exchange $J^{int}=-5\times 10^{-4}\textrm{J}/\textrm{m}^2$; the shape anisotropy fields were calculated using demagnetizing factors extracted from OOMMF simulations. Panels (b) and (d) show the corresponding calculated hysteresis loops.  In panel (a), filled symbols are used for ascending field (P$\rightarrow$ AP$\rightarrow$ SF) and open symbols for descending field (SF$\rightarrow$ AP$\rightarrow$ P).}
\end{figure}

\begin{figure}
\includegraphics[width=8.5cm]{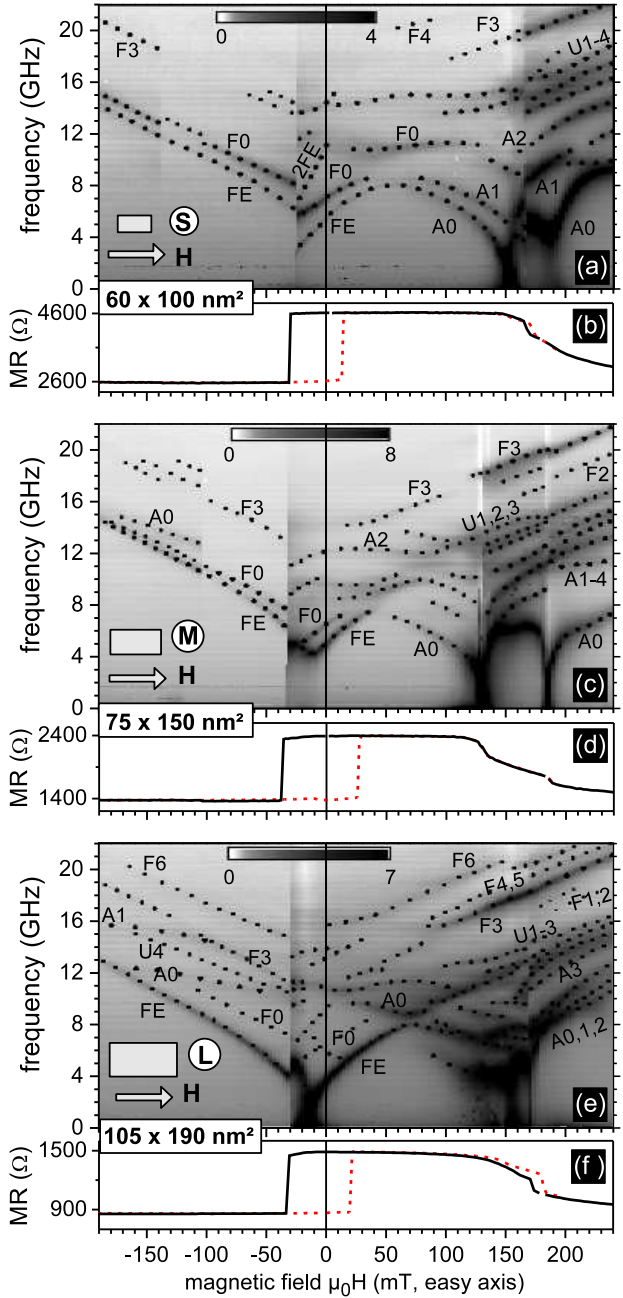}
\caption{\label{fig2} Power spectrum density (PSD, log scale) versus descending (SF$\rightarrow$AP$\rightarrow$P) easy axis applied field for a device of (a) size S, (c) size M, and (e) size L. Panels (b), (d), and (f): corresponding hysteresis loops.}
\end{figure}
%
%

%%%%%%%%%%%%%%%%%%%%%%%%%%%%%%%%%%%%%%%%%%%%%%%%%%%%%%%%%%%%%%%%%%%%%%%%%%
%
%
\begin{figure}
\includegraphics[width=8.5cm]{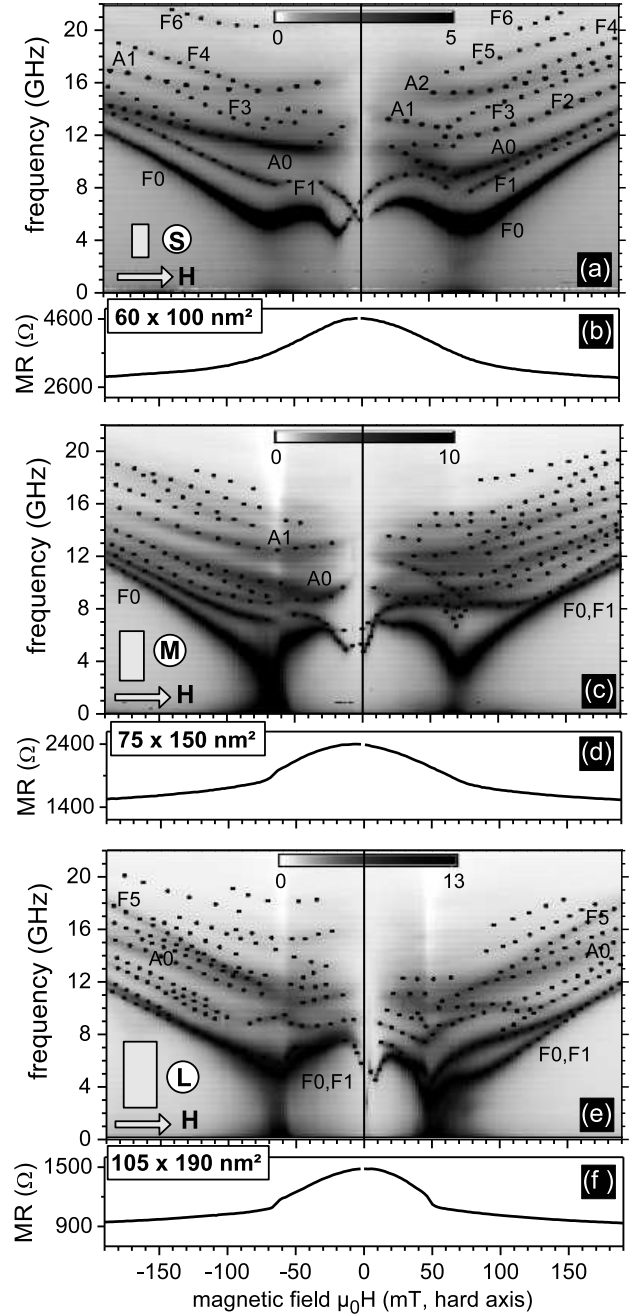}
\caption{\label{fig3}Power spectrum density (PSD, log scale) versus (ascending) hard axis field for the same devices as in Fig.~\ref{fig2}: (a)~size S, (c)~size M, and (e)~size L. Panels (b),~(d),~(f): corresponding hysteresis loops.}
\end{figure}

In this section, we describe the characteristics of the spectra measured for easy axis and hard axis applied fields as well as their dependence on the pillar size. Note that there is no device possessing simultaneously on EA and HA all the properties stated as typical of a particular pillar size. The properties described in the text are therefore those observed on a majority of the EA spectra and a majority of the HA spectra, but not always for the same devices.
For the basic identification of the observed spin wave modes we recall in Fig.~\ref{fig1}(a), (c) the mode dispersion for a nanopillar consisting of a SAF and an ideal free layer (i.e. the latter is assumed not to interact with the SAF) in macrospin approximation. 
EA and HA spectrum each contain two types of modes: the uniform FL modes and the uniform acoustic mode of the SAF. The acoustic SAF mode is thereby the lower of the two SAF eigenmodes and corresponds to oscillations, for which the in-plane components of the SAF layer magnetizations stay antiparallel, i.e. the dynamical magnetizations oscillate $180^{\circ}$ out-of-phase. The high-frequency second eigenmode of the SAF, the optical mode (in-plane components of dynamical magnetizations in-phase), is not detected in the measured frequency range up to 26 GHz, and will therefore not be mentioned further.

\subsection{Eigenexcitations for easy axis applied field}
\label{sec:resultEA}

\subsubsection{Size independent properties}

For all pillar sizes, EA spectra (Fig.~\ref{fig2}) contain two groups of modes. The first group consists of V-shaped modes basically symmetric about zero field for high positive and negative fields and history dependent in the hysteretic field region with a discontinuity at the coercive field. Since this is the typical behavior of FL modes (Fig.~\ref{fig1}(a),(b)), the modes in this group are labeled with F. The second mode group consists of modes having a minimum at or near the spin-flop field of the SAF. This being the characteristics of the acoustic modes of the SAF (Fig.~\ref{fig1}(a),(b)), the modes in this group are labeled with A. Modes that cannot be assigned unambiguously to one of the groups are labeled with U (like unidentified). Within each group the modes are numbered consecutively with increasing frequency. The lowest FL mode FE has been labeled differently, because it shows in several aspects a qualitatively different behavior than the next higher mode F0. As we shall see in the following subsections, there is evidence that it might belong to excitations localized near the layer edges (hence the second label E). Occasionally, second harmonics, such as the mode 2FE in Fig.~\ref{fig2}(a), are observed.\\

There are three FL modes common to all pillar sizes: the modes F0 and FE, visible in the P and low-field AP state, and - with a size invariant spacing of $5-6$ GHz to F0 - the mode F3, mainly visible at high fields.\\
In the SF region at high positive fields, several SAF modes are observed, the intense lowest modes having either one single minimum (Fig.~\ref{fig2}(e)) or two minima at different fields (Fig.~\ref{fig2}(a), (c)). The minima are always positioned in the vicinity of significant changes in the slope of the resistance in the hysteresis loop. The occurrence of more than one such slope change clearly indicates that the magnetization of the reference layer does not undergo a single abrupt transition as associated with the SF, but that there are several domains with different transition fields. 
%In the case of a single slope change, two possible scenario have been observed: If the slope change is discontinuous, the positions of the minimum or minima in the SAF modes do not coincide with those of the resistance jump, but the SAF modes rather show a discontinuity at this field value (cf. Figs.~\ref{fig2}(a), (e)). If, on the other hand, the slope change is continuous, a single minimum is observed at the field of the slope change (similar to the situation at $+130$ mT in Fig.~\ref{fig2}(c)). Empirically, in the former case, the resistance was found to be hysteretic at the SF, whereas in the latter case, it was not. 
Independent of the number of observed minima at the SF, the SAF modes show complicated, irregular structures, indicating strong non-uniformities of the SAF layer magnetizations in this field region, and will therefore not be discussed in more detail. \\

The modes after the SF, which are labeled with U, comprise the lowest FL modes and higher order SAF modes. They cannot be identified with certainty, because both mode types have similar frequency, slope, and intensity in this field region. Moreover, above the SF, the large angle between the FL magnetization and the RL magnetization boosts the experimental sensitivity to both FL and SAF modes (see section \ref{sec:sensitivity}) causing in particular for sizes S and M abrupt changes in the mode intensity, such that FL modes suddenly appearing above the SF field may be misinterpreted as SAF modes. \\
%For some specific devices (of any pillar size) both FE and F0 undergo an abrupt increase of the intensity at very high negative field (in Fig.~\ref{fig2}(a) at -140 mT, in Fig.~\ref{fig2}(c) at  -105 mT), while the resistance increases simultaneously by 10-20 Ohm (not visible on the scale of Figs.~\ref{fig2}(b),(d)). Occasionally, this process is accompanied by the appearance of one or two SAF modes (A0 and A1 in Fig.~\ref{fig2}(c)). The underlying reason  is very likely a change in the magnetic configuration of the RL (the field being too high for changes in the FL): its magnetization, or parts of it, must have turned away from the EA, probably in the course of the second SF transition of the SAF. Like for the SF at positive field discussed above, the intensity change is caused by the finite angle between (parts of) the RL magnetization and the FL magnetization.\\
For some devices, a change of mode intensity is also observed at a high negative field value (cf. F0, F3 in Fig.~\ref{fig2}(a) at -140 mT, or A0 in Fig.~\ref{fig2}(c) at -105 mT), at which the resistance changes by 10-20 Ohm (not visible on the scale of Figs.~\ref{fig2}(b),(d)). The reason for this is a change in the micromagnetic configuration, very likely of the RL (see section \ref{sec:sensitivity}).\\

Finally, in the AP state, gaps with pillar size dependent opening from 2 to 0.5~GHz are observed in the modes FE and, if visible, F0 (see e.g. Figs.~\ref{fig2}(e) and (a)). We will see that this is a consequence of the mutual dipolar coupling between the FL and the SAF leading to anticrossing of FL and SAF modes.

\subsubsection{Size dependent properties}

In the following, we describe the size specific  properties of the EA spectra. We will see that with increasing pillar size, the spectra undergo characteristic changes, some of which are caused directly by the increasing dimensions, while others are most likely consequences of an increasing non-uniformity of the magnetization.\\
 
{\bf Size S}\\
In the EA spectra of the smallest pillars (Fig.~\ref{fig2}(a)) the modes FE, F0, and F3 are mostly the only observed FL modes. The opening of the anticrossing gap in F0 is typically 2 GHz. F3 has for all devices a frequency of $22-23$ GHz at $+240$ mT. FE and F0 show a slightly larger device-to-device variation: For some devices, the mode FE is visible up to very high negative fields (as in Fig.~\ref{fig2}(a)) with a roughly constant spacing to F0 of about 1.5 GHz, for others it is observed in the low-field region only where the resistance departs from its saturation value, i.e. where the (FL) magnetization shows signs of increasing non-uniformity; in this case, FE rapidly approaches F0 for increasing field and vanishes, once the resistance has reached its saturation value. This correlation of FE to the non-uniformity of the static magnetization suggests that FE might be an edge mode. Another observation in favor of this supposition is that the mode FE has a noticeably higher slope than F0 (due to its  tendency to approach F0 asymptotically from below), which would not be the case if both modes were volume modes. 
The average frequency of F0 is about 16 GHz at $-190$ mT and 6.5 GHz at zero field.\\
%This assumption is corroborated by the fact that at zero field, FE shows a much higher and qualitatively different sensitivity to small HA fields than F0. On the other hand, the high-field visibility of FE for the other devices without their showing any sign of enhanced non-uniformity is obviously in contradiction to FE being an edge mode.  Whatever the nature of the mode FE, there seems to be a correlation between its visibility and the frequency of F0: for the 3 devices with low-field visibility of FE, F0 has a frequency of about 16 GHz at $-190$ mT and 6.5 GHz at zero field, whereas for the 2 devices with high-field visibility of FE, F0 is shifted to higher frequencies by about 1 GHz. Note that the device in Fig.~\ref{fig2}(a) happens to have a $15\%$ smaller FL magnetization than the other devices of size S, resulting in a lower slope of the FL modes and a reduced coercivity.\\

{\bf Size M}\\
The EA spectra of devices of pillar size M (Fig.~\ref{fig2}(c)) differ from those of the size-S devices in the following points: The mode FE has developed a minimum at low fields, which for some devices may almost reach zero frequency. This effect is likely to be caused by the increased non-uniformity of the magnetization in this field region. Occasionally, the mode F0, too, becomes deformed, though much less than FE. The relative intensity of F0 typically decreases, whereas that of FE increases. The observed evolution of FE with the pillar size, too, corroborates the assumption that this mode might be an edge mode. The frequencies of all modes decrease typically  by 1 or 2 GHz at high fields. 
%The frequency of F0 has decreased by 2 GHz at high fields, that of F3 by 1 GHz; at low fields, the frequency of F0 has practically remained unchanged, whereas the frequency of FE is basically the same as for size S; at $-190$ mT, it is typically 14 GHz, such that F0 and FE approach each other for increasing negative field. 
The gap opening in the modes FE and F0 has decreased to 1-1.5 GHz. At very high fields, additional modes, most likely belonging to the FL, appear just above or below the mode F3.\\

{\bf Size L}\\
For the size-L devices (Fig.~\ref{fig2}(e)), the  gap opening in the mode FE does not exceed 0.5-1~GHz, and the frequencies of FE and F3 have decreased by another GHz. The spacing between FE and F3 has not changed with respect to size M or S. Above F3 three additional FL modes (F4 to F6) with a spacing of 1~GHz, and below F3 two extremely weak modes (F1, F2) have appeared. As for size M, the mode FE is strongly deformed in the low field region and has still gained intensity with respect to F0. 
% As an obvious non-uniformity effect, a very weak, hook-shaped additional mode is sometimes observed near the minimum of FE.

\subsection{Eigenexcitations for hard axis applied field}
\label{sec:resultHA}

\subsubsection{Size independent properties}

In Fig.~\ref{fig3} are depicted the HA spectra of the same devices as in Fig.~\ref{fig2}. The FL modes have a characteristic W-shape with two minima in the lowest modes at about $\pm70$~mT. Though 70~mT is for all pillar sizes substantially higher than the measured anisotropy fields, the minima are often\cite{Counil:JAP:2005} interpreted as to correspond to the saturation of the free layer magnetization along the HA (cf. also Fig.~\ref{fig1}(c),(d)).\\
At or near zero field, the modes F0 and F1 show typically one, sometimes two sharp minima, which become deeper with increasing pillar size, and which are not present in the macrospin HA spectrum (Fig.~\ref{fig1}(c)). We therefore suspect that at least one of the modes observed at low fields is actually an edge mode FE.  As a matter of fact, if at zero EA field the lowest mode is - as we think - an edge mode, then the lowest mode at zero HA field must be an edge mode, too, because zero EA and zero HA field are formally identical. The observed field dependence of the frequency of the mode F0 then implies that its character must be changing (continuously) from edge mode at low fields to volume mode at high fields. Such a progressive change of the mode character would be consistent with the expected saturation process of the magnetization along the HA: for very low HA field, the magnetization in the central part of the layer is aligned along the EA due to the shape anisotropy, and only in narrow zones along the short edges of the rectangle the magnetization starts to align with the HA. For increasing field, these edge zones (domains) expand continuously towards the layer center, until for some field value the volume magnetization and finally the magnetization in the zones along the long edges of the layer saturate along the HA. The oscillations of this increasing part of the magnetization parallel to the field would obviously correspond to edge modes at low fields, and to volume modes at high fields; for intermediate field values, they would have a mixed character.\\
Finally, for most devices, we also observe the almost horizontal lowest acoustic modes of the SAF, A0 and A1 (see e.g. Fig. \ref{fig3}(a)), where A0 has a frequency of typically $10-12$~GHz at zero field. Note that in particular for size-S  devices the SAF modes are strongly asymmetric w.r.t. zero field, even though the FL modes and the hysteresis loop are basically symmetric. As will be shown in section~\ref{sec:extrapara}, this asymmetry cannot be explained by a misalignment of the external field with the HA. 

\subsubsection{Size dependent properties}
{\bf Size S}\\
The HA spectra of size-S devices are characterized by rounded saturation minima of the FL modes at $\pm 70$ mT (cf. Fig.~\ref{fig3}(a)). The frequency minima of the mode F0 are thereby not zero, but raised to values between 4 and 6~GHz, which is a consequence of the mutual dipolar coupling between the FL and the SAF, as we shall see in section~\ref{sec:discussion}.  
Typically 5 to 7 FL modes, F0 to F6 in Fig.~\ref{fig3}(a), are observed, where the first two modes have frequencies of 12 and 14~GHz, respectively, at $\pm190$~mT. The lowest mode F0 has much higher intensity than the other modes, and its frequency at high fields is the same for all measured devices of size S. The frequencies of the higher modes slightly vary from device to device. (In Fig.~\ref{fig3}(a) the modes F2 and A1 seem to accidently coincide at positive fields; however, for other devices,  F2 is clearly resolved.) 
The presence of F1 and F2 in the HA spectra for several devices of size S with rather high intensity is of great importance, because at least the mode F2 is not observed in the EA spectra, not even after the SF where the experimental sensitivity is comparable to that on the HA (see section~\ref{sec:sensitivity}).\\

{\bf Size M}\\
In the spectra of pillar size M (Fig.~\ref{fig3}(c)) the saturation minima at $\pm 70$ mT in the mode F0 are much deeper than for size S, which is consistent with the lower dipolar coupling between the FL and the SAF concluded from the EA hysteresis loops and spectra (see also section \ref{sec:discussion}). 
The minima at zero-field are considerably sharper than for size S indicating increasing importance of edge domain effects. The acoustic SAF modes have with about 10 GHz at zero field for A0 a slightly lower frequency than for size S, which is either due to a smaller interlayer exchange or the increasing non-uniformity of the SAF layer magnetizations (resulting e.g. on the EA in the observed stepwise switching in the SF region). Finally, the overall mode spacing has noticeably decreased compared to size S, as should be expected.\\

{\bf Size L}\\
In the spectra of pillar size L (Fig.~\ref{fig3}(e)) the minima in the modes F0 and F1 at $\pm 70$ mT reach, as for size M, markedly lower frequencies than for size S. Both F0 and F1 are strongly deformed in the vicinity of their minima and may even cross each other. The minimum at zero-field has still become slightly deeper, the impact of edge domains now being dominant. Contrary to size S, the shape of the modes in the low and medium field region is strongly device dependent and sensitive to small changes of the field direction.

%%%%%%%%%%%%%%%%%%%%%%%%%%%%%%%%%%%%%%%%%%%%%%%%%%%%%%%%%%%%%%%%%%%%%%%%%%%%%%%%%%%%%%%%%%%%%

\section{Outcomes and limits of the macrospin model}
\label{sec:kittel}

Before making a detailed and rigorous analysis of the field dependence of the modes frequencies in the next section, we start by attempting to model the free layer modes F0 using conventional Kittel fits. The aim is twofold: motivate the need for a more elaborate analysis by showing quantitative and qualitative limits of the macrospin approximation, and obtain approximate starting values for the magnetizations.\\

Approximating the free layer as an isolated rectangular platelet with only shape anisotropy, its ferromagnetic resonance frequency is described by the well-known Kittel law, which for EA applied field ($x$-direction) reads
$$\omega^2=\gamma_{0}^2 [H^{appl}+H_k][H^{appl}+(N^z-N^x)M_S],$$
and for HA field ($y$-direction) 
$$\omega^2=\gamma_{0}^2 [H^{appl}-H_k][H^{appl}+(N^z-N^y)M_S],$$
where $M_S$ is the saturation magnetization of the free layer and $H_k=(N^y-N^x)M_S$ the shape anisotropy field.\\ %Surface anisotropy is negligible in our pillars (see following section).
Applying Kittel fits to the modes F0 in the high field regions of the spectra in Figs.~\ref{fig2}-\ref{fig3}, allows us to extract 
$M_S$ and $H_k$ for the different pillar sizes, independently for EA and HA.
Using the demagnetizing factors $N^x$, $N^y$, and $N^z$ of Ref.~\onlinecite{Cornelissen:JAP:2009}, we obtain from the modes F0 in the EA spectra the following values for $M_S$ and $H_k$: for size S $\mu_0 M_S=1.14$~T (for most devices of size S: 1.3~T) and $\mu_0 H_k=37$~mT, for size M $\mu_0 M_S=1.04$~T and $\mu_0 H_k=35$~mT, and for size L $\mu_0 M_S=0.91$~T and $\mu_0 H_k=30$~mT. In comparison, for the modes FE, larger magnetizations (1.3 to 1.1~T), but much smaller anisotropy fields (less than 12~mT) are obtained.
Similarly, the modes F0 on the HA yield for size S $\mu_0 M_S=1.41$~T and $\mu_0 H_k=55$~mT (universal for size S), for size M $\mu_0 M_S=1.40$~T and $\mu_0 H_k=76$~mT, and for size L $\mu_0 M_S=1.41$~T and $\mu_0 H_k=78$~mT.\\

The minimum requirement for these values to be reasonable approximations is that the magnetizations and shape anisotropy fields extracted from EA and HA spectrum of the same device are roughly equal. However, as can be seen, both $M_S$ and $H_k$ are considerably larger on the HA, the discrepancies becoming larger with increasing pillar size. In addition, on the HA - and, if the mode FE is used, also on the EA - the anisotropy fields are neither consistent with the extracted magnetization nor with the anisotropies found by astroid measurements\cite{Cornelissen:JAP:2009} (cf. section \ref{sec:exp}).
Therefore, treating the free layer and the SAF as uncoupled systems consisting of uniformly magnetized layers is obviously insufficient to describe the eigenexcitations of nanopillars. The next section will be dedicated to a rigorous treatment of spin waves in a coupled three-layer system with lateral confinement.

\section{Model of spin wave modes in nanopillars}
\label{sec:model}
\subsection{Dipolar-exchange spin waves with quantized wavevectors}
\label{sec:model3L}

\subsubsection{Eigenexcitations of coupled three-layer system}

An MTJ nanopillar consists basically of three confined magnetic layers: the free layer, which will be labeled with the index ``F'', and below the two SAF layers - the reference (top) layer and the pinned (bottom) layer - labeled with indices ``1''  and ``2'', respectively. The magnetization dynamics in each layer $l\in \{F,1,2\}$ of this coupled three-layer system is governed by the Landau-Lifshitz equation.\\
For small amplitude precessions, the 
magnetization $\vec M_{l}(\vec r,t)$ can be decomposed in zeroth order approximation into a time  independent uniform (U) equilibrium component $\vec M_{l}^U$ (saturation magnetization $M_l$) and a small perpendicular dynamical part $\delta \vec M_{l}^U(\vec r,t)$. Static non-uniformities of the equilibrium magnetization will be discussed in section~\ref{sec:sensitivity}.
% $\vec M_{l}(\vec r,t)=\vec M_{l}+\delta \vec M_{l}(\vec r,t).$
The dynamical component  $\delta \vec M_{l}^U(\vec r,t)$ can be approximated as a sum of plane spin waves, 
\begin{equation}
\delta \vec M_{l}^U(\vec r,t)=\Re \sum_{\vec k}\delta \vec M_{l}^U(\vec k) {e^{i \vec k \vec r-i \omega_{\vec k} t}},\label{eq:deltaM}
\end{equation}
where the wavevectors $\vec k=(k_x,k_y,0)$ of the partial waves are quantized due to the spatial confinement of the layers. The out-of-plane component $k_z$ is zero for all modes in the experimental scope due to the very small layer thicknesses of $2-3~\textrm{nm}$. The quantization of the in-plane components $k_x,k_y$ will be discussed in detail later on. The frequencies $\omega_{\vec k}$ of the partial waves are the eigenfrequencies of the three-layer system.\\

In the effective fields acting on the magnetizations the following interactions have been taken into account:
the applied field ${\vec H}^{appl}$, the exchange bias field acting on the bottom layer of the SAF (coupling constant $J^{eb}$), the interlayer exchange coupling of the SAF layers (coupling constant $J^{int}$), and the (intralayer) exchange interaction in each layer (exchange stiffness constant $A_l$), as well as the demagnetizing fields and mutual dipolar coupling of the layers.
For the demagnetizing fields we use the standard tensor expression for uniformly magnetized ellipsoidal bodies, where the diagonal components of the diagonal (self-)demagnetizing tensors $\mathbf{N}_l$ are the demagnetizing factors $N_l^{x}$, $N_l^{y}$, $N_l^{z}$ of the rectangular layers. Although this approximation is expected to be satisfying for the static demagnetizing field, it is rather crude for the dynamical part, since the dynamical magnetization is non-uniform unless $\vec k=0$. \\
The fields resulting from mutual dipolar coupling are given by  analogous expressions where the (self-)demagnetizing tensors of trace 1 are replaced by the mutual demagnetizing tensors\cite{Newell:JGP:1993} $\mathbf{N}_{ml}$ of trace 0 ($l,m\in \{F,1,2\},\,l\not=m$). For the given pillar geometry, $\mathbf{N}_{ml}$ is diagonal, too, as can easily be shown using the formulae for the tensor components in Ref.~\onlinecite{Newell:JGP:1993}. The diagonal components will be referred to as the mutual dipolar coupling constants $N_{ml}^{x}$, $N_{ml}^{y}$, and $N_{ml}^{z}$. \\
Note that there is no significant perpendicular surface anisotropy at the top and bottom surfaces of the layers in MTJs, as has been demonstrated in Refs.~ \onlinecite{Cornelissen:JAP:2009,Kubota:APL:2006}. Since the impact of the bias current on the experimental spectra has been found to be negligible, we do not include current-based interactions, such as spin-torque or the Oersted field. The latter does e.g. not exceed 1~mT for a current of 0.3~mA and an impact diameter of 100~nm.\\

With these approximations the Landau-Lifshitz equations of the three pillar layers become a system of $3 \times 3=9$  coupled linear equations for the components of the dynamical magnetizations $\delta\vec M_{l}^U(\vec k)$. It can be solved as the eigenvalue problem of the $9\times9$ coefficient matrix $\mathbf {F}$ of the 9-component vector $(\delta\vec M_{F}^U(\vec k),\delta\vec M_{1}^U(\vec k),\delta\vec M_{2}^U(\vec k))$ describing the dynamics of the three-layer system as a whole. The eigenvalues of $\mathbf {F}$ are the eigenexcitations $\omega_{\vec k}$ of the three-layer system and can be calculated numerically as a function of the applied field, yielding the expected spin wave spectra $\omega_{\vec k}(H^{appl})$ of the nanopillar.

%%%%%%%%%%%%%%%%%%%%%%%%%%%%%%%%%%%%%%%%%%%%%%%%%%%%%%%%%%%%%%%%%%%%%%%%%%%%%%%%%%%%%%%%%

\subsubsection{Quantization of in-plane wavevector}

The in-plane components $k_x,~k_y$ of the wavevector are determined by the boundary conditions (BC) imposed on the dynamical magnetization~\eqref{eq:deltaM} at the lateral layer boundaries $x=\pm L_{x}/2$ and $y=\pm L_{y}/2$. For simplicity we will consider the $x$-component (along the long edge of the rectangle) as an example, where any of the following statements hold equally for the $y$-component with $x$ and $y$ permuted.\\
For the $x$-component the BC read:
\begin{equation}
\left[\frac{\partial}{\partial \xi_{x}} \delta \vec M_{l}^U(\xi_x,\xi_y)\pm d_{x}^{\pm}\,\delta \vec M_{l}^U(\xi_x,\xi_y)\right]_{\xi_{x}=\pm \frac{1}{2}}=0.\label{eq:BC}
\end{equation}
where $\xi_{x}={x}/L_{x}$. Eq. ~\eqref{eq:BC} is a modified version of the effective BC derived by Guslienko {\it et al.}\cite{Guslienko:PRB:2005} for thin magnetic stripes. In difference to Ref. \onlinecite{Guslienko:PRB:2005} we allow for different pinning parameters $d_{x}^{+}$ and $d_{x}^{-}$ at opposite boundaries $x=\pm L_{x}/2$ to account for potential asymmetries in the pinning expected from a real device. Moreover, instead of using the analytical expression (5) in Ref. \onlinecite{Guslienko:PRB:2005} to calculate the (dimensionless) pinning parameters, we will extract approximate values for $d_{x}^{\pm}$ from the experimental spectra (see section \ref{sec:extrapara}).\\

Applying the BC \eqref{eq:BC} to the sinusoidal mode profile
\begin{equation}
\Re {e^{i \vec k \vec r}}= \sin(k_{x} x+\phi_{x})\sin(k_{y} y+\phi_{y}), \label{eq:sines}
\end{equation}
of the partial spin waves in $\delta \vec M_{l}^U$~\eqref{eq:deltaM} yields for the wavevector component $k_{x}$ and the phase $\phi_{x}$ the quantization conditions 
%With the spatial dependence of $\delta \vec M_{l}^U$~\eqref{eq:deltaM} rewritten as 
%\begin{equation}
%\Re {e^{i \vec k \vec r}}= \sin(k_{x} x+\phi_{x})\sin(k_{y} y+\phi_{y}), \label{eq:sines}
%\end{equation}
%the BC \eqref{eq:BC} yield for the wavevector component $k_{x}$ and the phase $\phi_{x}$ the two conditions 
\begin{equation}
\mp k_{x}L_{x}\,\cot\left(\pm k_{x}\frac{L_{x}}{2}+\phi_{x}\right)=d_{x}^{\pm}.\label{eq:BCnew}
\end{equation}
It is convenient to express $k_{x} L_{x}$ in the argument of the cotangent as multiples of $\pi$, thus defining the - in general nonintegral - mode numbers 
\begin{equation}
n_{x} = \frac{k_{x} L_{x}}{\pi} \label{eq:defn}
\end{equation}
of the quantized spin wave modes $(n_x,n_y)$.\\
For symmetric pinning, $d_{x}^{+}=d_{x}^{-}=d_{x}$, it follows from \eqref{eq:BCnew} that the cotangent has to be antisymmetric, yielding $\phi_{x}^s=\pi/2$ or $\phi_{x}^a=0$, i.e. symmetric or antisymmetric wavefunctions \eqref{eq:sines}. In the limiting case of totally unpinned BC, $d_{x}=0$, the mode numbers $n_{x}^0$ are integers, starting at 0, and the corresponding wavefunctions alter between symmetric and antisymmetric for successive mode numbers, starting with symmetric, such that there are always antinodes at both boundaries.\\
For finite values $d_{x}>0$ of the pinning, the mode numbers $n_{x}$ are no longer integers. Plotting $n_{x}$ versus $d_{x}$ by means of eqs.~\eqref{eq:BCnew} and \eqref{eq:defn} shows that with increasing $d_{x}$, the deviations $\Delta n_{x}$ of $n_{x}$ from the corresponding integral values $n_{x}^0$ of the unpinned case increase continuously from $\Delta n_{x}=0$ for $d_{x}=0$ (unpinned) to $\Delta n_{x}=1$ for $d_{x}=\infty$ (totally pinned). Therefore, the mode numbers for total pinning, $n_{x}^{\infty}=n_{x}^0+1$, are integers again. For a fixed intermediate value $d_{x}$, the deviation $\Delta n_{x}$ of the mode number $n_{x}$ from the corresponding integral mode number $n_{x}^0$ is found to rapidly decrease with increasing $n_{x}^0$. For a given pinning, the mode numbers are therefore no independent variables: once one mode number (e.g. that of the lowest mode) has been fixed, all other mode numbers are fixed, too. \\  
In case of slightly asymmetric pinning, $d_{x}^{+}\not=d_{x}^{-}$, the phase $\phi_{x}$ differs from the values $\phi_{x}^{s,a}$ by a small phase shift $\Delta\phi_{x}$, such that the wavefunctions are no longer totally symmetric or antisymmetric. In this case, the mode numbers $n_{x}$ are necessarily non-integral. In the hypothetic case of totally asymmetric pinning, $d^{+}_{x}=0$ and $d^{-}_{x}=\infty$ (or vice versa), $\Delta n_{x}=0.5$ and $\Delta\phi_{x}=\Delta\phi_{x}^{max}=\pi/4$. For arbitrary pinning, $\Delta\phi_{x}$ is an unknown function of $d_{x}^{\pm}$ and $n_{x}$.\\

%The pinning is determined by two competing interactions: the dipolar interaction, dominating for large layer dimensions, favors pinned BC, whereas the exchange interaction, dominating for small layer dimensions, favors unpinned BC. 
The pinning for a given in-plane direction of a magnetic element depends on its dimensions and in addition on the inhomogeneity of the internal field.\cite{Bayer:book:2006,Guslienko:PRB:2005,Kostylev:054422} 
Consequently, the mode numbers are expected to be larger for the $x$-direction than for the $y$-direction of the same pillar, and possibly different for easy and hard axis applied field.

%%%%%%%%%%%%%%%%%%%%%%%%%%%%%%%%%%%%%%%%%%%%%%%%%%%%%%%%%%%%%%%%%%%%%%%%%%%%%%%%%%%%%%%%%%%%%%%%

\begin{table}

\begin{tabular}{p{7.5cm}}
\begin{tabular}{p{3.5cm} p{4.5cm}}
(a)&
\end{tabular}\\
\begin{tabular}{p{2.1cm} p{1.6cm}||p{1.3cm}|p{1.8cm}}
~~~~~~~~~~~~~~field& region &   $ {M_{l}^{V}}(\theta)$ & $\overline{\delta M_l^E}$ \\
\hline\hline
easy axis &  & &\\
\hline
below $2^{nd}$ SF   &  $H<H_{SF2}$ & $M_l\sin\theta_0$ & $ $\\&  & &\\
P state   &  high H & $M_l\Delta\theta$ & $ $\\
   & low H& $M_l\Delta\theta$ & $\overline{\delta M_l^E}$ \\&  & &\\
AP state    & low H & $M_l\Delta\theta$ & $\overline{\delta M_F^E}\,\delta_{lF}$\\
  &  high H & $M_l\Delta\theta$ & $\overline{\delta M_1^E}\,\delta_{l1}$\\& & & \\
above SF     & $H>H_{SF}$ & $M_l\sin\theta_0$ &\\
&  & \\
\hline\hline
hard axis & & &\\
\hline
         & $\lvert H\rvert>0$ &$M_l\sin\theta_0$ &\\
       &$ \,H\approx 0$ & $M_l\Delta\theta$ & $\overline{\delta M_l^E}$
\end{tabular}
\end{tabular}

\begin{tabular}{p{7.5cm}}
~~\\
\begin{tabular}{p{1.7cm} p{6cm}}
(b)&\\
&~~~~~~~~~~~~ integral ${\overline{W}_x}(n_{x},\phi_{x})$
\end{tabular}\\
\begin{tabular}{p{1.4cm}||p{2.3cm}|p{3.3cm}}
mode ~~~~~~ number~$n_{x}^0$&  for weak pinning $(\Delta n_{x}\ll1)$ & for strong pinning ~~~~~~~~~~~~~~~~~~~~~ $(\Delta n_{x}\approx 1)$\\
\hline\hline
0 & 1  & ${2}/{\pi}$\\
odd   & $\Delta\phi_{x}/n_{x}\cdot{2}/\pi$ & $\Delta\phi_{x}(1-\Delta n_{x})/{n_{x}}$ \\
even   & ${\Delta n_{x}}/{n_{x}}$ & $1/n_{x}\cdot2/\pi$ \\
\end{tabular}
\end{tabular}

%\begin{tabular}{p{7.5cm}}
%~~\\
%\begin{tabular}{p{1.7cm} p{6cm}}
%(c)&\\
%\end{tabular}\\
%\begin{tabular}{p{2.0cm}|p{1.7cm}||p{3.2cm}}
%mode numbers $(n_{x}^0,n_{y}^0)$& example modes& micromagnetic states, for which $\overline{\delta M_l^E}\not=0$\\
%\hline\hline
%(even, even) & (0,0), (2,0)  & S-state\\
%(even, odd)   & (0,1), (2,1)  & -- \\
%(odd, even)   & (1,0), (3,0)  & C-state \\
%(odd, odd)   & (1,1)  & S-, C-, and flower-state 
%\end{tabular}
%\end{tabular}
\caption{\label{tab:sensitivity} Dependence of the magneto-resistance noise \eqref{eq:MRnoise-princ} on the static micromagnetic configuration of layer $l\in\{F,1\}$ and the mode character. (a)~leading order contributions of volume magnetization, ${M_{l}^{V}}(\theta)$, and edge domain contributions, $\overline{\delta M_l^E}(n_{x},n_{y},\phi_{x},\phi_{y},\theta)$, versus easy and hard axis applied field in different field regions. $\delta_{lm}$ is the Kronecker symbol. (b)~leading order terms of integral ${\overline{W}_x}(n_{x},\phi_{x})$ of the wavefunction versus the mode number $n_{x}=n_{x}^0+\Delta n_{x}$ in the regime of weak and strong pinning. ${\overline{W}_y}$ is given by analogous expressions.  }
%(c)~dependence of edge-domain-induced mode visibility on micromagnetic state for weak pinning; for strong pinning, $\overline{\delta M_l^E}$ is negligible.

\end{table}

%%%%%%%%%%%%%%%%%%%%%%%%%%%%%%%%%%%%%%%%%%%%%%%%%%%%%%%%%%%%%%%%%%%

\subsection{Expected experimental sensitivity}

\label{sec:sensitivity}
\subsubsection{Formulation of the problem}
As described in section \ref{sec:exp}, the experimental spin wave spectra are obtained by measuring the voltage noise of the pillar. To be more precise, we measure the average of the local voltage noise over the pillar area. The local voltage noise is the product of the local current density and the local magneto-resistance (MR) noise generated by spin waves in the free layer (FL) and the reference layer (RL). In the ideal case of a homogeneous in-plane distribution of the current, the measured voltage noise is proportional to the average of the local MR noise. For the sake of simplicity, we will derive the expected MR noise for excitations in the FL, where the analogous expressions for the RL are obtained by permuting the indices ``F'' and ``1''. The consequences of inhomogeneities will be discussed later in this section.\\
The MR noise signature of a partial spin wave with wavevector $\vec k$ representing the FL mode $(n_x,n_y)$ is in linear order given by the square of
\begin{equation}
{\delta R_F(\vec k)} = \frac{1}{S_{pil}} \int_{S_{pil}} \vec M_1(\vec r)\cdot \delta {\vec M}_{F}^U(\vec k)\,\Re{e^{i \vec k \vec r}}\, d {\vec r}\label{eq:deltaR}
\end{equation}
where ${S_{pil}}$ is the pillar area, $\vec M_1(\vec r)$ the micromagnetic equilibrium magnetization of the RL, $\delta {\vec M}_{F}^U(\vec k)$ the amplitude vector and $\Re{e^{i \vec k \vec r}}$ the spatial dependence (wavefunction) of the spin wave in the FL (cf. eq.~\eqref{eq:deltaM}). 
Decomposing $\vec M_1(\vec r)$ as before into a uniform macrospin component ${\vec M}_1^U$, dominating in the central (volume) part of the layer, and the remaining $\vec r$-dependent edge domain components $\delta {\vec M}_1^E(\vec r)$,  yields as final expression for the MR variation~\eqref{eq:deltaR}
\begin{eqnarray}
\lefteqn{\delta R_F(n_{x},n_{y},\phi_{x},\phi_{y},\theta)}\nonumber\\
&=&\delta {M}_F^U(\vec k) \,\left[  {M_1^{V}}(\theta)\,{\overline{W}_x}(n_{x},\phi_{x})\,{\overline{W}_y}(n_{y},\phi_{y})\right.\nonumber\\
&&\phantom{\delta {M}_F^U(\vec k)\,}\left.+\overline{\delta M_l^E}(n_{x},n_{y},\phi_{x},\phi_{y},\theta)\right].\label{eq:MRnoise-princ}
\end{eqnarray}
The first term, ${M_1^{V}}\,{\overline{W}_x}\,{\overline{W}_y}$, is the contribution of the uniform volume magnetization to the MR noise (hence the superscript V). ${M_1^{V}}(\theta)=\sin \theta\,{M_1}$ is the projection of ${\vec M}_{1}^U$  onto $\delta {\vec M}_F^U(\vec k)$, where $\theta$ denotes the angle between the macrospins of FL and RL. $\overline{W}_x$ and $\overline{W}_y$ are the integrals of the $x$- and $y$-dependent factors of the wavefunction \eqref{eq:sines}, respectively.\\
The second term, $\overline{\delta M_1^E}$, is the contribution of the static edge domain magnetization components of the RL to the noise signature of the (volume) FL mode $(n_{x},n_{y})$ (not to be confused with edge modes). It is the spatial average of the projection of $\delta {\vec M}_1^E(\vec r)$ onto  $\delta {\vec M}_F^U(\vec k)$  weighted by the wavefunction \eqref{eq:sines}.\\
Mathematical expressions for $\overline{W}_x$ ($\overline{W}_y$) and $\overline{\delta M_1^E}$ as well as details on the derivation of eq.~\eqref{eq:MRnoise-princ} can be found in appendix~\ref{sec:appendix-sensitivity}. For the following discussion it is sufficient to consider the leading order terms of these quantities listed in Table~\ref{tab:sensitivity}. \\ 
The terms ${M_1^{V}}$ and $\overline{\delta M_1^E}$, resulting from the equilibrium magnetization, obviously depend on the static micromagnetic configuration of the pillar layers. The quantities $\overline{W}_x$, $\overline{W}_y$, and again
$\overline{\delta M_1^E}$, involving the wavefunction, depend on the symmetry properties of the spin wave mode. 
In the following, we will show which modes are expected to be observed in the experimental spectra under which conditions by analyzing the micromagnetic configuration and mode character dependence of the experimental sensitivity.

\subsubsection{Micromagnetic configuration dependence of sensitivity}

The micromagnetic configuration of the pillar is sensitive to direction and strength of the external field, which is why in the following we distinguish between easy and hard axis applied field and identify field regions of distinct values ${M_l^{V}}$ and $\overline {\delta {M}_l^E}$ ($l\in\{F,1\}$).\\
The largest contributions to the MR noise and hence highest sensitivity are expected in the field regions of maximum volume magnetization contributions ${M_l^{V}}(\theta)$, i.e. where  $\sin \theta\gg 0$. As can be seen from Table~\ref{tab:sensitivity}(a), 
a large zeroth order term $\sin \theta_0$ is found for non-zero HA fields, above the SF at positive EA fields and below the SF at negative EA fields. In the AP and P state, $\sin \theta_0=0$, such that the leading order terms are of first order in $\Delta\theta$. $\theta_0$ thereby denotes the angle between the FL and RL macrospins in case of an ideal pillar, and  $\Delta\theta$ a small deviation of $\theta$ from $\theta_0$ caused by a  misalignment of the exchange bias field or the external field  with the symmetry axes of the layers.\\
The presence of edge domains on the EA, as marked in Table~\ref{tab:sensitivity}(a), can be deduced from the hysteresis loops, as is explained in detail in  appendix~\ref{sec:appendixA-micro}. 
Whether these edge domains give non-zero contributions $\overline {\delta {M}_l^E}$ to the MR noise depends on the symmetry properties of both the micromagnetic state and the wavefunctions (see appendix~\ref{sec:appendixA-mode} for details). \\

In summary, high sensitivity to both FL and SAF modes can be expected on the HA at any finite field value and on the EA above the SF at positive fields. Weak higher modes will therefore be visible, if at all, in these field regions (cf. Figs~\ref{fig2},\ref{fig3}). The sensitivity below the SF at negative EA fields is also enhanced, though substantially less than for the other two cases. The change in intensity can be nicely seen in Fig.~\ref{fig2}(a) at -140 mT, and in Fig.~\ref{fig2}(c) at -105 mT.\\
For EA fields between the two SF fields, i.e. in AP and P state, modes become visible only through the misalignment $\Delta\theta$ of the macrospins or through edge domains of appropriate symmetry. From the latter, slightly increased sensitivity is expected for FL modes in the AP state at high positive fields just below the SF, and for both FL and SAF modes at low fields in P and AP state. The presence of edge domains may thereby entail the appearance of the corresponding edge modes in the spectra.

\subsubsection{Mode character dependence of sensitivity}

The quantities $\overline{W}_x$, $\overline{W}_y$, and $\overline{\delta M_1^E}$ depend on the symmetry properties of the wavefunction of the mode $(n_x,n_y)$. As the edge domain contributions become effective mainly in the low-field region, in which our model is in any case not expected to be accurate,
we consider only the integrals  $\overline{W}_x$ and $\overline{W}_y$ belonging to the volume magnetization contribution ${M_{l}^V}$, dominating at high fields.\\
In Table~\ref{tab:sensitivity}(b) the leading order terms of $\overline{W}_x$ are listed as a function of the mode number  $n_{x}=n_{x}^0+\Delta n_{x}$ in the regime of weak and strong pinning. $n_{x}^0$, $\Delta n_{x}$, and $\Delta \phi_{x}$ are thereby defined as in section~\ref{sec:model3L}. The $y$-dependent factor $\overline{W}_y$ is given by analogous expressions.\\
For zero pinning ($\Delta n_{x}=\Delta\phi_{x}=0$), the fundamental mode $n_{x}^0=0$ is the only visible mode. In the presence of pinning, the higher modes $n_{x}^0\geq1$ begin to appear: for symmetric pinning ($\Delta n_{x}>0,\,\Delta\phi_{x}=0$) only those with symmetric wavefunctions (even $n_{x}^0$), in case of asymmetric pinning ($\Delta n_{x}>0,\,\Delta\phi_{x}\not=0$) also those with antisymmetric wavefunctions (odd $n_{x}^0$).\\

For weak pinning ($\Delta n_{x}\ll1$), ${\overline{W}_x}$ is for all higher modes $n_{x}^0\geq1$ of first order in a small quantity: in $\Delta n_{x}$ for even $n_{x}^0$, in $\Delta\phi_{x}$ for odd $n_{x}^0$. In appendix~\ref{sec:appendix-sensitivity}, we show that the expected intensities of the  higher modes $(n_{x}^0,n_{y}^0)=(1,0),$  (0,1), and (2,0) are about two orders of magnitude lower than that of the fundamental mode $(n_{x}^0,n_{y}^0)=(0,0)$, whereas the mode $(n_{x}^0,n_{y}^0)=(1,1)$ is expected to have a four orders of magnitude lower intensity than (0,0).\\ 
%It follows that ${\overline{W}_x}(n_{x},\phi_{x})$ decreases rapidly with increasing mode number, firstly because of the factor $1/n_{x}$, secondly because $\Delta n_{x}$ becomes smaller as $n_{x}^0$ increases (cf. section~\ref{sec:model3L}).
For strong pinning ($\Delta n_{x}\approx1$), the natural reference mode numbers are the mode numbers $n_{x}^{\infty}=n_{x}^0+1$ of total pinning: even (odd) $n_{x}^0$ in the table correspond to odd (even) $n_{x}^{\infty}$. ${\overline{W}_x}$ for symmetric wavefunctions (even $n_{x}^0$, odd  $n_{x}^{\infty}$) has now become a zeroth order quantity like for the lowest mode $n_{x}^0=0$ ($n_{x}^{\infty}=1$), whereas for antisymmetric wavefunctions (odd $n_{x}^0$, even  $n_{x}^{\infty}$) it has become second order in $(1-\Delta n_{x})$ and $\Delta\phi_{x}$. Consequently, the lowest higher order modes close to $(n_{x}^{\infty},n_{y}^{\infty})=(3,1)$ and (1,3) will have intensities comparable to that of the fundamental mode $(n_{x}^{\infty},n_{y}^{\infty})=(1,1)$. Modes with an even mode number $n_{x,y}^{\infty}$ are expected to be at least four orders of magnitude weaker than (1,1). \\
%For all modes, ${\overline{W}_x}(n_{x},\phi_{x})$ decreases with increasing mode number as $1/n_{x}$; the decrease of $\Delta n_{x}$ and $\Delta\phi_{x}$, however, does not affect modes with even $n_{x}^0$, and is partly compensated for odd $n_{x}^0$, as the factor $(1-\Delta n_{x})$ increases with decreasing $\Delta n_{x}$.

The above results on the expected relative mode intensity have been obtained under the assumption of homogeneous current density and homogeneous saturation magnetizations. Under these conditions the voltage noise is proportional to the MR noise, and - at high fields where the edge domain contributions are negligible - the MR noise is proportional to $({\overline{W}_x}{\overline{W}_y})^2$. As we have seen, in this case higher modes become visible if the integrals $\overline{W}_x$ and $\overline{W}_y$ are non-zero, that is for non-zero asymmetric pinning. However, even in the absence of pinning, the measured voltage noise can be non-zero, namely if the saturation magnetization or the current distribution are inhomogeneous, because then the average~\eqref{eq:deltaR} over the pillar area becomes an integral of a generally unharmonic - and for asymmetric inhomogeneities also asymmetric - function. For real devices, we may therefore expect finite sensitivity to most of the higher modes.

\section{Extraction of material parameters}
\label{sec:extrapara}

%%%%%%%%%%%%%%%%%%%%%%%%%%%%%%%%%%%%%%%%%%%%%%%%%%%%%%%%%%%%%%%%%%%%%%%%%%%%%%%%%%
  
\begin{table}
\begin{tabular}{p{2.4cm}|p{2.4cm}|p{3.0cm}}
    parameter  & & extracted value\\
\hline\hline
      saturation& $\mu_0 M_{F,1}$&$1.27~ \textrm{T}$\\
      magnetization&$\mu_0 M_2$ &$1.4~ \textrm{T}$\\
      \hline
      exchange&$A_{F,1}$ & $18.0 \times 10^{-12}\textrm{J}/\textrm{m}$\\
      stiffness&$A_2$ &$14.0 \times 10^{-12}\textrm{J}/\textrm{m}$\\
      \hline
      exchange bias  &$J^{eb}$ & $4.5 \times 10^{-4}\textrm{J}/\textrm{m}^2$\\
      interlayer exch. &$J^{int}$ & $-3.9 \times 10^{-4}\textrm{J}/\textrm{m}^2$\\
      
\hline\hline
      lateral pillar    &$L_x$ & 100~\textrm{nm}\\
      dimensions& $L_y$&60~\textrm{nm}\\
      \hline
      demagnetizing&$(N_F^{x},N_F^{y},N_F^{z})$ &$(0.035,0.065,0.9)$\\
      factors& $(N_{1,2}^{x},N_{1,2}^{y},N_{1,2}^{z})$&$(0.027,0.049,0.924)$\\
      \hline
      dipolar coupling& $(N_{F1}^{x},N_{F1}^{y},N_{F1}^{z})$&$(0.01,0.018,-0.028)$\\
       constants&$(N_{F2}^{x},N_{F2}^{y},N_{F2}^{z})$ &$(0.005,0.009,-0.014)$\\
       &$(N_{12}^{x},N_{12}^{y},N_{12}^{z})$ &$(0.007,0.012,-0.019)$\\
       \hline
       mode numbers  & $(n_x,n_y)$&f00, a00: $(0.4,0.0)$\\
       easy axis& &f10, a10: $(1.13,0.0)$\\
       & &f01, a01: $(0.4,1.0)$\\
         & &f20, a20: $(2.05,0.0)$\\
         & &f11, a11: $(1.13,1.0)$\\
       \hline
       mode numbers  & $(n_x,n_y)$&f00, a00: $(0.2,0.0)$\\
       hard axis& &f10, a10: $(1.05,0.0)$\\
       & &f01, a01: $(0.2,1.0)$\\
         & & f20, a20: $(2.02,0.0)$\\
         & & f11, a11: $(1.05,1.0)$\\
         & & f21, a21: $(2.02,1.0)$
\end{tabular}
\caption{\label{tab:para} Material and geometry parameters used to calculate the spectra and hysteresis loops in Fig.~\ref{fig4}. Error bars for the parameters are given in the text. The labels $fmn$ and $amn$ associated to the mode numbers are those in Fig.~\ref{fig4}. }
\end{table}

%**************************************************************************************

In section \ref{sec:model3L}, we have derived the mode frequencies $\omega_{\vec k}$ as a function of the material parameters $M_l$, $A_l$, $J^{eb}$, $J^{int}$, the geometry parameters $(L_x,L_y)$, $(N_l^{x}$, $N_l^{y}$, $N_l^{z})$, and $(N_{ml}^{x}$, $N_{ml}^{y}$, $N_{ml}^{z})$, as well as the mode numbers $(n_x,n_y)$. 
In this section, we will finally extract these parameters from the experimental spectra.
Since the model is based on the assumption of uniform equilibrium magnetizations, its application can be expected to be reasonable only for the smallest pillar size S, for which the non-uniformities of the magnetization had been found to be minimum. We assume that in the EA spectra, the modes F0 and F3, and in the HA spectra, the modes F0 to F5 and A0, A1 are, at sufficiently high fields, volume modes describable by the model.\\

\begin{figure}
\includegraphics[width=8.5cm]{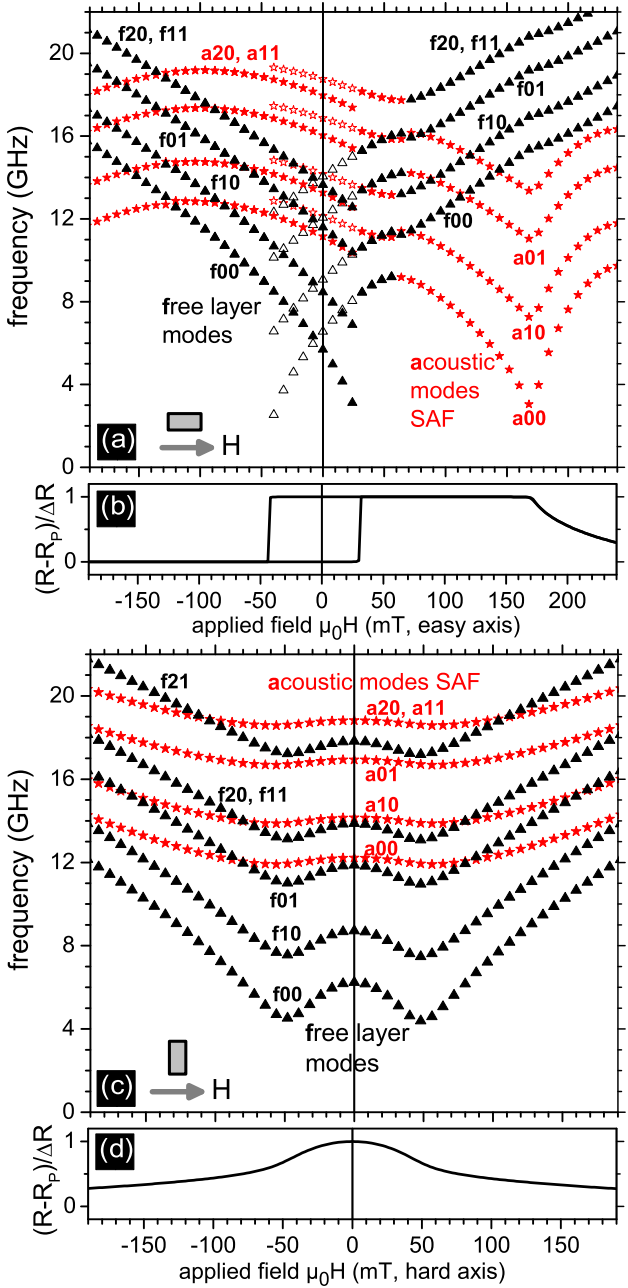}
\caption{\label{fig4}Calculated mode frequencies versus magnetic field along easy axis (a) and hard axis (c) for a pillar of size S. Panels (b) and (d) show the corresponding calculated hysteresis loops. The parameters used to calculate the spectra are given in Tab.~\ref{tab:para}.  The modes f20, f11 have practically identical frequency, which is why only one mode is displayed. In panel (a), filled symbols are used for ascending field (P$\rightarrow$ AP$\rightarrow$ SF) and open symbols for descending field.}
\end{figure}

%%%%%%%%%%%%%%%%%%%%%%%%%%%%%%%%%%%%%%%%%%%%%%%%%%%%%%%%%%%%%%%%%%%%%%%%%%%%%%%%%%%%%%

As a matter of fact, not all of the above quantities are free input parameters to the model. In
appendix~\ref{sec:appendix-extrapara} we show that on the basis of the measured layer dimensions, basic OOMMF simulations, previous works published in the literature, and a couple of reasonable assumptions, the number of free parameters can be reduced to the following quantities: the three mutual dipolar coupling constants $N_{F1}^{x}$, $N_{F2}^{x}$, $N_{12}^{x}$; magnetization $M_F$ and exchange stiffness constant $A_F$ of the free layer; interlayer exchange $J^{int}$ and exchange bias $J^{eb}$ of the SAF; the mode numbers of the lowest FL and SAF mode. \\
% Literature thin film and bulk values of the material parameters are given in appendix~\ref{sec:appendix-literature}.
In appendix~\ref{sec:appendix-extrapara}, we extract minimum and maximum values for the remaining parameters by adjusting the calculated modes and hysteresis loops to the corresponding experimental data (Figs.~\ref{fig2},\ref{fig3}(a),(b)). In particular, we show that from the experimental constraints it follows that the pinning of the magnetization at the boundaries must be weak. \\
Best overall agreement of experiment and theory is obtained for the parameters in Tab.~\ref{tab:para}. The calculated spectra and hysteresis loops are shown in Fig.~\ref{fig4}. In the following, we will point out similarities as well as differences between experiment and theory, and will discuss the ``technical'' reliability of the parameter values. The physical consequences will be discussed in the next section. \\

The HA spectra (Figs.~\ref{fig3}(a),\ref{fig4}(c)) show quantitative agreement for the FL modes F0 to F3 at medium and high fields, and for the lowest SAF mode A0 at high negative field only. However, there is no calculated mode corresponding to the mode F4; the frequency of the mode f21, which is the next higher mode after f20/f11, is much too high for F4.  
A possible reason for this discrepancy might be our for higher modes rather crude approximation of the dynamical demagnetizing field (see below).
The systematic asymmetry of the SAF modes, resulting in a much lower frequency of A0 at positive field, cannot be accounted for by a tilting of the field w.r.t. to the HA, as this  would affect both FL and SAF modes as well as the hysteresis loop. It might rather be caused by a misalignment of the exchange bias field, determining the magnetic symmetry axis of the SAF, with the geometrical symmetry axes of the rectangle, coinciding with the magnetic symmetry axes of the free layer.\\ %Note that asymmetries in the pillar shape, being more or less random, would not explain the systematic character of the asymmetry.
In the zero-field region, the differences between calculated and experimental HA spectra become substantial, as the measured modes develop pronounced minima, whereas the model predicts a local frequency maximum.\\
On the EA (Figs.~\ref{fig2},\ref{fig4}(a)), the calculated modes f00 and f20/f11 fit the {\it average} experimental modes F0 and F3 (see section~\ref{sec:resultEA}) rather well. In particular, the gap opening in the mode F0 is reproduced in the theoretical spectrum. 
The modes f01 (F2), and - if none of the unidentified modes U is F1- also f10, are not observed in the experimental EA spectra. Their absence is likely to be due to either an overall lack of intensity, first noticeable for the weakest modes, or a lower sensitivity to these particular modes on the EA, although our considerations in section~\ref{sec:sensitivity} yield no satisfying explanation for the different visibility for EA (above the spin-flop) and HA field under the made assumptions. 
For the SAF modes there can be only qualitative agreement due to the multiple SF in the experiment, which - like the low-frequency supposed edge modes FE - can of course not be described by a macrospin-based model.\\
Finally, the calculated hysteresis loops in Fig.~\ref{fig4}(b),~(d) are in qualitative agreement with the measured loops.\\

Main error sources for any of the parameters are obviously the various assumptions in section~\ref{sec:model3L} and appendix~\ref{sec:appendix-extrapara}. Additional uncertainties come from insufficient experimental data, e.g. the ignorance of the actual SF field or device-to-device diversity of mode frequencies. In particular, our representation of the dynamical demagnetizing field in the standard  tensor expression for uniformly magnetized ellipsoidal bodies can be expected to be a reasonable approximation only for the lowest mode, because there the dynamical magnetization is indeed almost uniform. As we extract the material parameters mainly from this mode, their values are only little affected by this approximation. The largest discrepancies are expected for the exchange stiffnesses, as they rely necessarily on the higher modes.\\
The value of $\mu_0 M_F$ is found to be between 1.25~T and 1.3~T , $A_F$ is expected to lie in the interval $(18\pm3) \times 10^{-12}\textrm{J}/\textrm{m}$. Under the assumptions in appendix~\ref{sec:appendix-extrapara} the same holds true for $M_1$ and $A_1$. The error of $M_2$ and $A_2$ is significantly larger than for the FL and RL, because of the additional dependence on the thin film value of the CoFe layer and the ignorance of the experimental exchange bias field. $\mu_0 M_2$ is expected to be contained in the interval $(1.4\pm0.1)$~T, and $A_2$  in $(16\pm4) \times 10^{-12}\textrm{J}/\textrm{m}$. $J^{int}$ is estimated to be $(-4.0\pm0.4)\times 10^{-4}\textrm{J}/\textrm{m}^2$, and $J^{eb}$ $(4.2\pm0.7)\times 10^{-4}\textrm{J}/\textrm{m}^2$.\\
The mode numbers of F0 (f00) have to be smaller than $(0.6,0.0)$, or $(0.3,0.2)$ for $n_y>0$, in order to ensure satisfactory agreement in frequency and a reasonable value for $A_F$. The agreement is better, if the mode numbers are chosen smaller on the HA than on the EA, and $n_x>n_y$. For the sake of simplicity, we have therefore set $n_y=0$.\\ 
$N_{F1}^{x}=0.01$ and $N_{F2}^{x}=0.005$ are uniquely determined by the experimental constraints with a maximum deviation of $\pm0.002$. The deviations of $\pm10$~nm of the lateral dimensions $L_x, L_y$ from the mean values will change all geometry related parameters accordingly.

\section{Discussion}

\label{sec:discussion}

In the previous sections, we have modeled the spin wave spectra of MTJ nanopillars as eigenexcitations of a coupled three-layer system with lateral confinement. In this section, we will see, which properties of the experimental spectra can be explained in the scope of this analytical model, and which cannot.
First, we will discuss the material parameters of the pillar extracted from the high field regions of the spin wave spectra. Thereafter, the low-field anomaly of the spectra and its relevance for applications will be discussed. Finally, we will summarize the properties of the experimental spectra, which are beyond the approximations of our model, including the pillar size dependence.

%%%%%%%%%%%%%%%%%%%%%%%%%%%%%%%%%%%%%%%%%%%%%%%%%%%%%%%%%%%%%%%%%%%%%%%%%%%%%%%%%%%%%%%%%
\subsection{Material and geometry parameters}

In this subsection, we will discuss the physical relevance of the extracted parameter values of the pillar in Table~\ref{tab:para}. \\

%\subsubsection{Magnetizations}

With $1.27$~T the saturation magnetization of the $\textrm{Co}_{60}\textrm{Fe}_{20}\textrm{B}_{20}$ layers of the pillar is significantly  reduced compared to the thin film value of $(1.8\pm0.1)$~T,\cite{Bilzer:JAP:2006, Cornelissen:JAP:2009} or the bulk value for the underlying $\textrm{Co}_{75}\textrm{Fe}_{25}$ of $(2.2\pm0.1)$~T.\cite{Bozorth:book:1993,Lamy:JAP:2005} 
% Note that this value of $M_S$ is valid for both field orientations, in contrast to the Kittel fits in section~\ref{sec:kittel}, where different values had been obtained on easy and hard axis. 
A reduction of the magnetization in nanopillars has already been observed in previous  studies on pillar devices.\cite{Kiselev:PRL:2004,Braganca:APL:2005,Petit:PRL:2007} 
Three scenarios are usually suggested to account for this phenomenon: process-induced damages,\cite{Krivorotov:Science:2005, Cornelissen:JAP:2009} current-induced heating,\cite{Lee:APL:2008,Devolder:JAP:2009} or a nonlinear change of the frequency with high mode amplitude.\cite{Mistral:APL:2006} As we work with low bias current, current induced heating can be excluded in our case. Similarly, since spin-torque induced auto-oscillations in our samples occur typically for currents above 1.6 mA for size L,\cite{Cornelissen:EPL:2009} high amplitude nonlinear effects as possible cause can be rejected, too. Therefore, some sort of process damage, such as ion implantation, diffusion, intrinsic chemical modifications or interface effects, must be at the origin of the magnetization reduction, whose further investigation exceeds the scope of this paper. \\

%\subsubsection{Exchange stiffness and boundary conditions}

Concerning the boundary conditions and exchange stiffness we have come to the following conclusions: Strong pinning can be ruled out in our pillars (see section~\ref{sec:extrapara}); reasonable agreement between calculated and experimental data is obtained under the assumption of weak pinning.  The pinning parameter deduced from the extracted mode numbers is with $d\leq1$ about 10 times smaller than the one calculated by means of eq.~(5) in Ref.~\onlinecite{Guslienko:PRB:2005} ($d\approx 10$) when using the material parameters of Tab.~\ref{tab:para}. Any value of $d$ substantially larger than 3 is found to yield mode numbers for the lowest mode very close to 1, i.e. strong pinning. This discrepancy between our result and the predictions of Guslienko's analytical model\cite{Guslienko:PRB:2005} are not understood, as the latter is expected to be valid in the regime of element thicknesses smaller than the exchange length as well.\\
We emphasize that, just as the magnetization, the exchange stiffness of the free layer does not exceed $2/3$ of the thin film value, independent of the boundary conditions. Therefore, the magnetic properties of the nanopillar can by no means be described by the values measured on the unprocessed thin films.\\

%\subsubsection{Dipolar coupling in tensor approximation}

The mutual dipolar coupling accounts for several features of the experimental spectra: In the HA spectra, the mutual dipolar coupling of the FL and the SAF raises the frequency minima of F0 by several GHz, pushes them to slightly higher fields, and lowers the slope of the modes, reducing their frequency at $\pm190$~mT by about 1~GHz. It also causes the bell shape of the HA hysteresis loop, by forcing the pillar into the AP state at low fields, and smoothes out the sharp bends at the anisotropy fields, which are  observed in the case of an uncoupled free layer. 
\\In the EA spectra, the gap openings in the mode F0 stem from the anticrossing of F0 with the acoustic SAF modes due to coupling-induced mode hybridization. Finally, the net dipolar coupling field created by the SAF layers and favoring the antiparallel configuration of the pillar causes a shift of the EA hysteresis loops to negative fields of 5~mT, which is approximately $50\%$ of the observed total shift. The remaining $50\%$ may be due to an unequal reduction of the coercive fields at positive and negative field, which occurs if the micromagnetic configuration causes the FL to switch more easily from the P state to the AP state, than from the AP state to the P state. Indeed, the FL magnetization is expected to be more non-uniform - and consequently easier to switch - in the low-field P state because of the mutual dipolar coupling field pointing antiparallel to the magnetizations in the P state, but parallel in the AP state.  \\
Within the diagonal tensor approximation of the mutual dipolar coupling, the tensor components are found to be significantly smaller than the values predicted by the formalism developed by Newell et al.\cite{Newell:JGP:1993} or by the simplified version using for the in-plane components of the mutual dipolar coupling tensor the corresponding components of the self-demagnetizing tensor, as is commonly practiced when modeling flip-flop switching in MRAM cells.\cite{Worledge:APL:2004b, Worledge:APL:2007} The coupling between e.g. the free layer and the reference layer of our pillars would be overestimated by Ref.~\onlinecite{Newell:JGP:1993} by a factor of 2, and by Refs.~\onlinecite{Worledge:APL:2004, Worledge:APL:2004b} by a factor of 3. 
A possible explanation for this reduction of the interlayer dipolar coupling may be that the coupling field extracted from the experiment is actually an {\it effective} mutual dipolar coupling field comprising the dipolar coupling due to the charges at the lateral layer boundaries as well as some N\'{e}el-type coupling resulting from the correlated roughness of the three magnetic layers. This orange-peel coupling may partially compensate the antiparallel coupling due to the charges at the layer edges. Another possibility is a reduction of the dipolar coupling due to the non-uniformity of the micromagnetic magnetization at the layer edges, though this effect should be small at high fields.\\ 
 
Finally, the extracted exchange bias energy and the interlayer exchange coupling are consistent with the large body of dedicated literature (see e.g. the values in appendix~\ref{sec:appendix-literature}).

%%%%%%%%%%%%%%%%%%%%%%%%%%%%%%%%%%%%%%%%%%%%%%%%%%%%%%%%%%%%%%%%%%%%%%%%%%%%%%%%%%%%%%%%%

\subsection{Low-field behavior and its relevance for applications}

In sections~\ref{sec:exp} and~\ref{sec:results}, we have seen that at low fields both EA and HA spectra show for all three pillar sizes unmistakable signs of non-uniform magnetizations: in the HA spectra, the FL modes possess at zero field, instead of the local maxima predicted by the model, sharp minima, whose depth increases with increasing pillar size, indicating increasing non-uniformity of the magnetization. In fact, the modes F0, F1 are likely to change character from volume modes at high and medium fields to edge modes at low field. The EA spectra contain low-frequency supposed edge modes FE, which become progressively deformed around zero field for increasing pillar size, i.e. for increasing non-uniformity of the magnetization. \\ % (It is interesting to note that the minimum at zero HA field is not observed in the otherwise very similar HA spectrum of the $200\times500\times15~\textrm{nm}^3$ single-layer ellipses  in Ref.~\onlinecite{Montoncello:PRB:2007}, in spite of the larger layer dimensions.) 
The non-uniformities of the magnetizations are expected to influence the switching dynamics of the pillar. The first consequence is that they lower the coercive field (as discussed in the previous paragraph), thus enlarging its difference to the shape anisotropy field. This effect has indeed been found to be particularly strong for size L (see section~\ref{sec:exp}).   
More importantly, the fact that the lowest mode is not the uniform mode, but an edge mode, will affect the magnetization reversal path in current-induced switching, favoring non-uniform reversal paths, as has already been concluded indirectly from reversal speed experiments.\cite{Devolder:PRL:2008}

%%%%%%%%%%%%%%%%%%%%%%%%%%%%%%%%%%%%%%%%%%%%%%%%%%%%%%%%%%%%%%%%%%%%%%%%%%%%%%%%
\subsection{Spin wave phenomena beyond the analytical approximations}

Based on the assumption of macrospin equilibrium magnetizations, our model is certain not to describe any effect resulting from non-uniformities of the magnetization. This is the case e.g. for the low-field behavior discussed in the previous paragraph, or the occurrence of more than one spin-flop transition of the SAF at positive EA field.\\
However, even for high fields and pillar size S where the model is expected to work reasonably well,  there are qualitative discrepancies between calculated and experimental spectra in frequency or visibility of higher order modes. \\
Similarly, the high-field evolution of the spin wave spectra with the pillar size is not consistent with the predictions by the model. Although the model allows to reproduce  qualitatively  the EA spectra for pillar size L under reasonable assumptions, it fails for the high-field HA spectra.

\section{Conclusions}

In this paper, we have studied the magnetic field dependence of the mode frequency of thermally excited spin waves in rectangular shaped MgO-MTJ nanopillars of different lateral sizes. The spin wave spectra (frequency versus easy and hard axis applied field) of individual devices were obtained using spectrally resolved electrical noise power measurements.\\
In all spectra, several independent quantized spin wave modes stemming from eigenexcitations in the free layer and the SAF layers of the MTJ have been observed. By diagonalizing the dynamical matrix of a system of three coupled, spatially confined magnetic layers, we have modeled the mode frequencies for the smallest pillar size, $60~\times~100~\textrm{nm}^2$, obtaining quantitative agreement for a majority of modes at high and medium applied fields. Our ability to detect a particular spin wave mode depends on the static micromagnetic configuration of the layers as well as on the symmetry properties of the mode. With the help of these discrimination criteria, we could identify the observed modes and extract the material parameters of the pillar (Tab.~\ref{tab:para}). The magnetizations and exchange stiffness constants were found to be significantly reduced compared to the corresponding thin film values, whereas the interlayer exchange coupling and the exchange bias are  consistent with their thin film counterparts. The interlayer dipolar coupling between the different layers could be well described in terms of an effective mutual dipolar coupling. Moreover, we could infer that the pinning of the magnetizations at the lateral boundaries must be weak.\\
Finally, at low fields and for larger pillar sizes, there is clear evidence for strong non-uniformities of the layer magnetizations, leading to qualitative differences between calculated and measured spin wave frequencies.

\begin{acknowledgments}
We thank Singulus Technologies A.G. for the layer deposition in a Timaris PVD system. A.~H. is supported by the European Community (EC) under the 6th FP for the Marie Curie RTN SPINSWITCH, contract no. MRTN-CT-2006-035327. The work in Leuven was supported by the EC program IST STREP, under contract no. IST-016939 TUNAMOS; S.~C. acknowledges IWT Flanders for financial support.
\end{acknowledgments}

%\bibliography{bibomain}

\begin{thebibliography}{40}
\expandafter\ifx\csname natexlab\endcsname\relax\def\natexlab#1{#1}\fi
\expandafter\ifx\csname bibnamefont\endcsname\relax
  \def\bibnamefont#1{#1}\fi
\expandafter\ifx\csname bibfnamefont\endcsname\relax
  \def\bibfnamefont#1{#1}\fi
\expandafter\ifx\csname citenamefont\endcsname\relax
  \def\citenamefont#1{#1}\fi
\expandafter\ifx\csname url\endcsname\relax
  \def\url#1{\texttt{#1}}\fi
\expandafter\ifx\csname urlprefix\endcsname\relax\def\urlprefix{URL }\fi
\providecommand{\bibinfo}[2]{#2}
\providecommand{\eprint}[2][]{\url{#2}}

\bibitem[{\citenamefont{Katine and Fullerton}(2008)}]{Katine:JMMM:2008}
\bibinfo{author}{\bibfnamefont{J.}~\bibnamefont{Katine}} \bibnamefont{and}
  \bibinfo{author}{\bibfnamefont{E.~E.} \bibnamefont{Fullerton}},
  \bibinfo{journal}{J. Magn. Magn. Mater.}
  \textbf{\bibinfo{volume}{320}}, \bibinfo{pages}{1217 }
  (\bibinfo{year}{2008}), ISSN \bibinfo{issn}{0304-8853}.

\bibitem[{\citenamefont{Chappert et~al.}(2007)\citenamefont{Chappert, Fert, and
  Van~Dau}}]{Chappert:Nature:2007}
\bibinfo{author}{\bibfnamefont{C.}~\bibnamefont{Chappert}},
  \bibinfo{author}{\bibfnamefont{A.}~\bibnamefont{Fert}}, \bibnamefont{and}
  \bibinfo{author}{\bibfnamefont{F.~N.} \bibnamefont{Van~Dau}},
  \bibinfo{journal}{Nat. Mater.} \textbf{\bibinfo{volume}{6}},
  \bibinfo{pages}{813} (\bibinfo{year}{2007}).

\bibitem[{\citenamefont{Deac et~al.}(2008)\citenamefont{Deac, Fukushima,
  Kubota, Maehara, Suzuki, Yuasa, Nagamine, Tsunekawa, Djayaprawira, and
  Watanabe}}]{Deac:Nature:2008}
\bibinfo{author}{\bibfnamefont{A.~M.} \bibnamefont{Deac}},
  \bibinfo{author}{\bibfnamefont{A.}~\bibnamefont{Fukushima}},
  \bibinfo{author}{\bibfnamefont{H.}~\bibnamefont{Kubota}},
  \bibinfo{author}{\bibfnamefont{H.}~\bibnamefont{Maehara}},
  \bibinfo{author}{\bibfnamefont{Y.}~\bibnamefont{Suzuki}},
  \bibinfo{author}{\bibfnamefont{S.}~\bibnamefont{Yuasa}},
  \bibinfo{author}{\bibfnamefont{Y.}~\bibnamefont{Nagamine}},
  \bibinfo{author}{\bibfnamefont{K.}~\bibnamefont{Tsunekawa}},
  \bibinfo{author}{\bibfnamefont{D.~D.} \bibnamefont{Djayaprawira}},
  \bibnamefont{and} \bibinfo{author}{\bibfnamefont{N.}~\bibnamefont{Watanabe}},
  \bibinfo{journal}{Nat. Phys.} \textbf{\bibinfo{volume}{4}},
  \bibinfo{pages}{803} (\bibinfo{year}{2008}), ISSN \bibinfo{issn}{1745-2473}.

\bibitem[{\citenamefont{Bayer et~al.}(2006)\citenamefont{Bayer, Jorzick,
  Demokritov, Slavin, Guslienko, Berkov, Gorn, Kostylev, and
  Hillebrands}}]{Bayer:book:2006}
\bibinfo{author}{\bibfnamefont{C.}~\bibnamefont{Bayer}},
  \bibinfo{author}{\bibfnamefont{J.}~\bibnamefont{Jorzick}},
  \bibinfo{author}{\bibfnamefont{S.~O.} \bibnamefont{Demokritov}},
  \bibinfo{author}{\bibfnamefont{A.~N.} \bibnamefont{Slavin}},
  \bibinfo{author}{\bibfnamefont{K.~Y.} \bibnamefont{Guslienko}},
  \bibinfo{author}{\bibfnamefont{D.~V.} \bibnamefont{Berkov}},
  \bibinfo{author}{\bibfnamefont{N.~L.} \bibnamefont{Gorn}},
  \bibinfo{author}{\bibfnamefont{M.~P.} \bibnamefont{Kostylev}},
  \bibnamefont{and}
  \bibinfo{author}{\bibfnamefont{B.}~\bibnamefont{Hillebrands}}, in
  \emph{\bibinfo{booktitle}{Spin Dynamics in Confined Magnetic Structures III}}
  (\bibinfo{year}{2006}), vol. \bibinfo{volume}{101} of
  \emph{\bibinfo{series}{Topics in Applied Physics}}, pp.
  \bibinfo{pages}{57--103}.

\bibitem[{\citenamefont{Bailleul et~al.}(2006)\citenamefont{Bailleul,
  H\"{o}llinger, and Fermon}}]{Bailleul:PRB:2006}
\bibinfo{author}{\bibfnamefont{M.}~\bibnamefont{Bailleul}},
  \bibinfo{author}{\bibfnamefont{R.}~\bibnamefont{H\"{o}llinger}},
  \bibnamefont{and} \bibinfo{author}{\bibfnamefont{C.}~\bibnamefont{Fermon}},
  \bibinfo{journal}{Phys. Rev. B}
  \textbf{\bibinfo{volume}{73}}, \bibinfo{eid}{104424}
  (pages~\bibinfo{numpages}{14}) (\bibinfo{year}{2006}).

\bibitem[{\citenamefont{Gubbiotti
  et~al.}(2007{\natexlab{a}})\citenamefont{Gubbiotti, Madami, Tacchi, Carlotti,
  Adeyeye, Goolaup, Singh, and Slavin}}]{Gubbiotti:JMMM:2007}
\bibinfo{author}{\bibfnamefont{G.}~\bibnamefont{Gubbiotti}},
  \bibinfo{author}{\bibfnamefont{M.}~\bibnamefont{Madami}},
  \bibinfo{author}{\bibfnamefont{S.}~\bibnamefont{Tacchi}},
  \bibinfo{author}{\bibfnamefont{G.}~\bibnamefont{Carlotti}},
  \bibinfo{author}{\bibfnamefont{A.}~\bibnamefont{Adeyeye}},
  \bibinfo{author}{\bibfnamefont{S.}~\bibnamefont{Goolaup}},
  \bibinfo{author}{\bibfnamefont{N.}~\bibnamefont{Singh}}, \bibnamefont{and}
  \bibinfo{author}{\bibfnamefont{A.}~\bibnamefont{Slavin}},
  \bibinfo{journal}{J. Magn. Magn. Mater.}
  \textbf{\bibinfo{volume}{316}}, \bibinfo{pages}{e338 }
  (\bibinfo{year}{2007}{\natexlab{a}}), ISSN \bibinfo{issn}{0304-8853}.

\bibitem[{\citenamefont{Gubbiotti et~al.}(2005)\citenamefont{Gubbiotti,
  Carlotti, Okuno, Grimsditch, Giovannini, Montoncello, and
  Nizzoli}}]{Gubbiotti:PRB:2005}
\bibinfo{author}{\bibfnamefont{G.}~\bibnamefont{Gubbiotti}},
  \bibinfo{author}{\bibfnamefont{G.}~\bibnamefont{Carlotti}},
  \bibinfo{author}{\bibfnamefont{T.}~\bibnamefont{Okuno}},
  \bibinfo{author}{\bibfnamefont{M.}~\bibnamefont{Grimsditch}},
  \bibinfo{author}{\bibfnamefont{L.}~\bibnamefont{Giovannini}},
  \bibinfo{author}{\bibfnamefont{F.}~\bibnamefont{Montoncello}},
  \bibnamefont{and} \bibinfo{author}{\bibfnamefont{F.}~\bibnamefont{Nizzoli}},
  \bibinfo{journal}{Phys. Rev. B} \textbf{\bibinfo{volume}{72}},
  \bibinfo{pages}{184419} (\bibinfo{year}{2005}).

\bibitem[{\citenamefont{Montoncello et~al.}(2007)\citenamefont{Montoncello,
  Giovannini, Nizzoli, Vavassori, Grimsditch, Ono, Gubbiotti, Tacchi, and
  Carlotti}}]{Montoncello:PRB:2007}
\bibinfo{author}{\bibfnamefont{F.}~\bibnamefont{Montoncello}},
  \bibinfo{author}{\bibfnamefont{L.}~\bibnamefont{Giovannini}},
  \bibinfo{author}{\bibfnamefont{F.}~\bibnamefont{Nizzoli}},
  \bibinfo{author}{\bibfnamefont{P.}~\bibnamefont{Vavassori}},
  \bibinfo{author}{\bibfnamefont{M.}~\bibnamefont{Grimsditch}},
  \bibinfo{author}{\bibfnamefont{T.}~\bibnamefont{Ono}},
  \bibinfo{author}{\bibfnamefont{G.}~\bibnamefont{Gubbiotti}},
  \bibinfo{author}{\bibfnamefont{S.}~\bibnamefont{Tacchi}}, \bibnamefont{and}
  \bibinfo{author}{\bibfnamefont{G.}~\bibnamefont{Carlotti}},
  \bibinfo{journal}{Phys. Rev. B}
  \textbf{\bibinfo{volume}{76}}, \bibinfo{eid}{024426}
  (pages~\bibinfo{numpages}{6}) (\bibinfo{year}{2007}).

\bibitem[{\citenamefont{Guslienko and Slavin}(2005)}]{Guslienko:PRB:2005}
\bibinfo{author}{\bibfnamefont{K.~Y.} \bibnamefont{Guslienko}}
  \bibnamefont{and} \bibinfo{author}{\bibfnamefont{A.~N.}
  \bibnamefont{Slavin}}, \bibinfo{journal}{Phys. Rev. B}
  \textbf{\bibinfo{volume}{72}}, \bibinfo{eid}{014463}
  (pages~\bibinfo{numpages}{5}) (\bibinfo{year}{2005}).

\bibitem[{\citenamefont{Vavassori
  et~al.}(2008{\natexlab{a}})\citenamefont{Vavassori, Bonanni, Busato, Bisero,
  Gubbiotti, Adeyeye, Goolaup, Singh, Spezzani, and
  Sacchi}}]{Vavassori:JPDApplPhys:2008}
\bibinfo{author}{\bibfnamefont{P.}~\bibnamefont{Vavassori}},
  \bibinfo{author}{\bibfnamefont{V.}~\bibnamefont{Bonanni}},
  \bibinfo{author}{\bibfnamefont{A.}~\bibnamefont{Busato}},
  \bibinfo{author}{\bibfnamefont{D.}~\bibnamefont{Bisero}},
  \bibinfo{author}{\bibfnamefont{G.}~\bibnamefont{Gubbiotti}},
  \bibinfo{author}{\bibfnamefont{A.~O.} \bibnamefont{Adeyeye}},
  \bibinfo{author}{\bibfnamefont{S.}~\bibnamefont{Goolaup}},
  \bibinfo{author}{\bibfnamefont{N.}~\bibnamefont{Singh}},
  \bibinfo{author}{\bibfnamefont{C.}~\bibnamefont{Spezzani}}, \bibnamefont{and}
  \bibinfo{author}{\bibfnamefont{M.}~\bibnamefont{Sacchi}},
  \bibinfo{journal}{J. Phys. D}
  \textbf{\bibinfo{volume}{41}}, \bibinfo{pages}{134014 (5pp)}
  (\bibinfo{year}{2008}{\natexlab{a}}).

\bibitem[{\citenamefont{Gubbiotti et~al.}(2006)\citenamefont{Gubbiotti, Madami,
  Tacchi, Carlotti, and Okuno}}]{Gubbiotti:PRB:2006}
\bibinfo{author}{\bibfnamefont{G.}~\bibnamefont{Gubbiotti}},
  \bibinfo{author}{\bibfnamefont{M.}~\bibnamefont{Madami}},
  \bibinfo{author}{\bibfnamefont{S.}~\bibnamefont{Tacchi}},
  \bibinfo{author}{\bibfnamefont{G.}~\bibnamefont{Carlotti}}, \bibnamefont{and}
  \bibinfo{author}{\bibfnamefont{T.}~\bibnamefont{Okuno}},
  \bibinfo{journal}{Phys. Rev. B}
  \textbf{\bibinfo{volume}{73}}, \bibinfo{eid}{144430}
  (pages~\bibinfo{numpages}{6}) (\bibinfo{year}{2006}).

\bibitem[{\citenamefont{Gubbiotti
  et~al.}(2007{\natexlab{b}})\citenamefont{Gubbiotti, Madami, Tacchi, Socino,
  Carlotti, and Ono}}]{Gubbiotti:MMMINT:2007}
\bibinfo{author}{\bibfnamefont{G.}~\bibnamefont{Gubbiotti}},
  \bibinfo{author}{\bibfnamefont{M.}~\bibnamefont{Madami}},
  \bibinfo{author}{\bibfnamefont{S.}~\bibnamefont{Tacchi}},
  \bibinfo{author}{\bibfnamefont{G.}~\bibnamefont{Socino}},
  \bibinfo{author}{\bibfnamefont{G.}~\bibnamefont{Carlotti}}, \bibnamefont{and}
  \bibinfo{author}{\bibfnamefont{T.}~\bibnamefont{Ono}}, \bibinfo{journal}{J. Appl. Phys.} \textbf{\bibinfo{volume}{101}},
  \bibinfo{eid}{09F502} (pages~\bibinfo{numpages}{3})
  (\bibinfo{year}{2007}{\natexlab{b}}).

\bibitem[{\citenamefont{Vavassori
  et~al.}(2008{\natexlab{b}})\citenamefont{Vavassori, Bonanni, Busato,
  Gubbiotti, Madami, Adeyeye, Goolaup, Singh, Spezzani, and
  Sacchi}}]{Vavassori:JAP:2008}
\bibinfo{author}{\bibfnamefont{P.}~\bibnamefont{Vavassori}},
  \bibinfo{author}{\bibfnamefont{V.}~\bibnamefont{Bonanni}},
  \bibinfo{author}{\bibfnamefont{A.}~\bibnamefont{Busato}},
  \bibinfo{author}{\bibfnamefont{G.}~\bibnamefont{Gubbiotti}},
  \bibinfo{author}{\bibfnamefont{M.}~\bibnamefont{Madami}},
  \bibinfo{author}{\bibfnamefont{A.~O.} \bibnamefont{Adeyeye}},
  \bibinfo{author}{\bibfnamefont{S.}~\bibnamefont{Goolaup}},
  \bibinfo{author}{\bibfnamefont{N.}~\bibnamefont{Singh}},
  \bibinfo{author}{\bibfnamefont{C.}~\bibnamefont{Spezzani}}, \bibnamefont{and}
  \bibinfo{author}{\bibfnamefont{M.}~\bibnamefont{Sacchi}},
  \bibinfo{journal}{J. Appl. Phys.} \textbf{\bibinfo{volume}{103}},
  \bibinfo{eid}{07C512} (pages~\bibinfo{numpages}{3})
  (\bibinfo{year}{2008}{\natexlab{b}}).

\bibitem[{\citenamefont{Bilzer et~al.}(2006)\citenamefont{Bilzer, Devolder,
  Kim, Counil, Chappert, Cardoso, and Freitas}}]{Bilzer:JAP:2006}
\bibinfo{author}{\bibfnamefont{C.}~\bibnamefont{Bilzer}},
  \bibinfo{author}{\bibfnamefont{T.}~\bibnamefont{Devolder}},
  \bibinfo{author}{\bibfnamefont{J.-V.} \bibnamefont{Kim}},
  \bibinfo{author}{\bibfnamefont{G.}~\bibnamefont{Counil}},
  \bibinfo{author}{\bibfnamefont{C.}~\bibnamefont{Chappert}},
  \bibinfo{author}{\bibfnamefont{S.}~\bibnamefont{Cardoso}}, \bibnamefont{and}
  \bibinfo{author}{\bibfnamefont{P.~P.} \bibnamefont{Freitas}},
  \bibinfo{journal}{J. Appl. Phys.} \textbf{\bibinfo{volume}{100}},
  \bibinfo{eid}{053903} (pages~\bibinfo{numpages}{4}) (\bibinfo{year}{2006}).

\bibitem[{\citenamefont{Cornelissen
  et~al.}(2009{\natexlab{a}})\citenamefont{Cornelissen, Bianchini, Helmer,
  Devolder, Kim, de~Beeck, Roy, Lagae, and Chappert}}]{Cornelissen:JAP:2009}
\bibinfo{author}{\bibfnamefont{S.}~\bibnamefont{Cornelissen}},
  \bibinfo{author}{\bibfnamefont{L.}~\bibnamefont{Bianchini}},
  \bibinfo{author}{\bibfnamefont{A.}~\bibnamefont{Helmer}},
  \bibinfo{author}{\bibfnamefont{T.}~\bibnamefont{Devolder}},
  \bibinfo{author}{\bibfnamefont{J.-V.} \bibnamefont{Kim}},
  \bibinfo{author}{\bibfnamefont{M.~O.} \bibnamefont{de~Beeck}},
  \bibinfo{author}{\bibfnamefont{W.~V.} \bibnamefont{Roy}},
  \bibinfo{author}{\bibfnamefont{L.}~\bibnamefont{Lagae}}, \bibnamefont{and}
  \bibinfo{author}{\bibfnamefont{C.}~\bibnamefont{Chappert}},
  \bibinfo{journal}{J. Appl. Phys.} \textbf{\bibinfo{volume}{105}},
  \bibinfo{eid}{07B903} (pages~\bibinfo{numpages}{3})
  (\bibinfo{year}{2009}{\natexlab{a}}).

\bibitem[{\citenamefont{Devolder
  et~al.}(2008{\natexlab{a}})\citenamefont{Devolder, Hayakawa, Ito, Takahashi,
  Ikeda, Katine, Carey, Crozat, Kim, Chappert et~al.}}]{Devolder:JAP:2008}
\bibinfo{author}{\bibfnamefont{T.}~\bibnamefont{Devolder}},
  \bibinfo{author}{\bibfnamefont{J.}~\bibnamefont{Hayakawa}},
  \bibinfo{author}{\bibfnamefont{K.}~\bibnamefont{Ito}},
  \bibinfo{author}{\bibfnamefont{H.}~\bibnamefont{Takahashi}},
  \bibinfo{author}{\bibfnamefont{S.}~\bibnamefont{Ikeda}},
  \bibinfo{author}{\bibfnamefont{J.~A.} \bibnamefont{Katine}},
  \bibinfo{author}{\bibfnamefont{M.~J.} \bibnamefont{Carey}},
  \bibinfo{author}{\bibfnamefont{P.}~\bibnamefont{Crozat}},
  \bibinfo{author}{\bibfnamefont{J.~V.} \bibnamefont{Kim}},
  \bibinfo{author}{\bibfnamefont{C.}~\bibnamefont{Chappert}},
  \bibnamefont{et~al.}, \bibinfo{journal}{J. Appl. Phys.}
  \textbf{\bibinfo{volume}{103}}, \bibinfo{eid}{07A723}
  (\bibinfo{year}{2008}{\natexlab{a}}).

\bibitem[{\citenamefont{Devolder et~al.}(2005)\citenamefont{Devolder, Crozat,
  Chappert, Miltat, Tulapurkar, Suzuki, and Yagami}}]{Devolder:PRB:2005}
\bibinfo{author}{\bibfnamefont{T.}~\bibnamefont{Devolder}},
  \bibinfo{author}{\bibfnamefont{P.}~\bibnamefont{Crozat}},
  \bibinfo{author}{\bibfnamefont{C.}~\bibnamefont{Chappert}},
  \bibinfo{author}{\bibfnamefont{J.}~\bibnamefont{Miltat}},
  \bibinfo{author}{\bibfnamefont{A.}~\bibnamefont{Tulapurkar}},
  \bibinfo{author}{\bibfnamefont{Y.}~\bibnamefont{Suzuki}}, \bibnamefont{and}
  \bibinfo{author}{\bibfnamefont{K.}~\bibnamefont{Yagami}},
  \bibinfo{journal}{Phys. Rev. B} \textbf{\bibinfo{volume}{71}},
  \bibinfo{eid}{184401} (pages~\bibinfo{numpages}{6}) (\bibinfo{year}{2005}).

\bibitem[{\citenamefont{Smith}(2001)}]{Smith:JAP:2001}
\bibinfo{author}{\bibfnamefont{N.}~\bibnamefont{Smith}}, \bibinfo{journal}{J.
  Appl. Phys.} \textbf{\bibinfo{volume}{90}}, \bibinfo{pages}{5768}
  (\bibinfo{year}{2001}).

\bibitem[{\citenamefont{Cornelissen
  et~al.}(2009{\natexlab{b}})\citenamefont{Cornelissen, Bianchini, Hrkac,
  de~Beeck, Lagae, Kim, Devolder, Crozat, Chappert, and
  Schrefl}}]{Cornelissen:EPL:2009}
\bibinfo{author}{\bibfnamefont{S.}~\bibnamefont{Cornelissen}},
  \bibinfo{author}{\bibfnamefont{L.}~\bibnamefont{Bianchini}},
  \bibinfo{author}{\bibfnamefont{G.}~\bibnamefont{Hrkac}},
  \bibinfo{author}{\bibfnamefont{M.~O.} \bibnamefont{de~Beeck}},
  \bibinfo{author}{\bibfnamefont{L.}~\bibnamefont{Lagae}},
  \bibinfo{author}{\bibfnamefont{J.-V.} \bibnamefont{Kim}},
  \bibinfo{author}{\bibfnamefont{T.}~\bibnamefont{Devolder}},
  \bibinfo{author}{\bibfnamefont{P.}~\bibnamefont{Crozat}},
  \bibinfo{author}{\bibfnamefont{C.}~\bibnamefont{Chappert}}, \bibnamefont{and}
  \bibinfo{author}{\bibfnamefont{T.}~\bibnamefont{Schrefl}},
  \bibinfo{journal}{Eur. Phys. Lett.} \textbf{\bibinfo{volume}{87}},
  \bibinfo{pages}{57001 (4pp)} (\bibinfo{year}{2009}{\natexlab{b}}).

\bibitem[{\citenamefont{Counil et~al.}(2005)\citenamefont{Counil, Kim,
  Devolder, Crozat, Chappert, and Cebollada}}]{Counil:JAP:2005}
\bibinfo{author}{\bibfnamefont{G.}~\bibnamefont{Counil}},
  \bibinfo{author}{\bibfnamefont{J.-V.} \bibnamefont{Kim}},
  \bibinfo{author}{\bibfnamefont{T.}~\bibnamefont{Devolder}},
  \bibinfo{author}{\bibfnamefont{P.}~\bibnamefont{Crozat}},
  \bibinfo{author}{\bibfnamefont{C.}~\bibnamefont{Chappert}}, \bibnamefont{and}
  \bibinfo{author}{\bibfnamefont{A.}~\bibnamefont{Cebollada}},
  \bibinfo{journal}{J. Appl. Phys.} \textbf{\bibinfo{volume}{98}},
  \bibinfo{eid}{023901} (pages~\bibinfo{numpages}{6}) (\bibinfo{year}{2005}).

\bibitem[{\citenamefont{Newell et~al.}(1993)\citenamefont{Newell, Williams, and
  Dunlop}}]{Newell:JGP:1993}
\bibinfo{author}{\bibfnamefont{A.~J.} \bibnamefont{Newell}},
  \bibinfo{author}{\bibfnamefont{W.}~\bibnamefont{Williams}}, \bibnamefont{and}
  \bibinfo{author}{\bibfnamefont{D.~J.} \bibnamefont{Dunlop}},
  \bibinfo{journal}{J. Geophys. Res.} \textbf{\bibinfo{volume}{98(B6)}},
  \bibinfo{pages}{9551} (\bibinfo{year}{1993}).

\bibitem[{\citenamefont{Kubota et~al.}(2006)\citenamefont{Kubota, Fukushima,
  Ootani, Yuasa, Ando, Maehara, Tsunekawa, Djayaprawira, Watanabe, and
  Suzuki}}]{Kubota:APL:2006}
\bibinfo{author}{\bibfnamefont{H.}~\bibnamefont{Kubota}},
  \bibinfo{author}{\bibfnamefont{A.}~\bibnamefont{Fukushima}},
  \bibinfo{author}{\bibfnamefont{Y.}~\bibnamefont{Ootani}},
  \bibinfo{author}{\bibfnamefont{S.}~\bibnamefont{Yuasa}},
  \bibinfo{author}{\bibfnamefont{K.}~\bibnamefont{Ando}},
  \bibinfo{author}{\bibfnamefont{H.}~\bibnamefont{Maehara}},
  \bibinfo{author}{\bibfnamefont{K.}~\bibnamefont{Tsunekawa}},
  \bibinfo{author}{\bibfnamefont{D.~D.} \bibnamefont{Djayaprawira}},
  \bibinfo{author}{\bibfnamefont{N.}~\bibnamefont{Watanabe}}, \bibnamefont{and}
  \bibinfo{author}{\bibfnamefont{Y.}~\bibnamefont{Suzuki}},
  \bibinfo{journal}{Appl. Phys. Lett.} \textbf{\bibinfo{volume}{89}},
  \bibinfo{eid}{032505} (pages~\bibinfo{numpages}{3}) (\bibinfo{year}{2006}).

\bibitem[{\citenamefont{Kostylev et~al.}(2007)\citenamefont{Kostylev,
  Gubbiotti, Hu, Carlotti, Ono, and Stamps}}]{Kostylev:054422}
\bibinfo{author}{\bibfnamefont{M.~P.} \bibnamefont{Kostylev}},
  \bibinfo{author}{\bibfnamefont{G.}~\bibnamefont{Gubbiotti}},
  \bibinfo{author}{\bibfnamefont{J.-G.} \bibnamefont{Hu}},
  \bibinfo{author}{\bibfnamefont{G.}~\bibnamefont{Carlotti}},
  \bibinfo{author}{\bibfnamefont{T.}~\bibnamefont{Ono}}, \bibnamefont{and}
  \bibinfo{author}{\bibfnamefont{R.~L.} \bibnamefont{Stamps}},
  \bibinfo{journal}{Phys. Rev. B} \textbf{\bibinfo{volume}{76}},
  \bibinfo{eid}{054422} (pages~\bibinfo{numpages}{8}) (\bibinfo{year}{2007}).

\bibitem[{\citenamefont{Bozorth}(1993)}]{Bozorth:book:1993}
\bibinfo{author}{\bibfnamefont{R.~M.} \bibnamefont{Bozorth}},
  \emph{\bibinfo{title}{Ferromagnetism}} (\bibinfo{publisher}{IEEE Inc., New
  York}, \bibinfo{year}{1993}).

\bibitem[{\citenamefont{Lamy and Viala}(2005)}]{Lamy:JAP:2005}
\bibinfo{author}{\bibfnamefont{Y.}~\bibnamefont{Lamy}} \bibnamefont{and}
  \bibinfo{author}{\bibfnamefont{B.}~\bibnamefont{Viala}},
  \bibinfo{journal}{J. Appl. Phys.}
  \textbf{\bibinfo{volume}{97}}, \bibinfo{eid}{10F910}
  (pages~\bibinfo{numpages}{3}) (\bibinfo{year}{2005}).

\bibitem[{\citenamefont{Rickart et~al.}(2005)\citenamefont{Rickart, Guedes,
  Negulescu, Ventura, Sousa, Diaz, MacKenzie, Chapman, and
  Freitas}}]{Rickart:EJPB:2005}
\bibinfo{author}{\bibfnamefont{M.}~\bibnamefont{Rickart}},
  \bibinfo{author}{\bibfnamefont{A.}~\bibnamefont{Guedes}},
  \bibinfo{author}{\bibfnamefont{B.}~\bibnamefont{Negulescu}},
  \bibinfo{author}{\bibfnamefont{J.}~\bibnamefont{Ventura}},
  \bibinfo{author}{\bibfnamefont{J.}~\bibnamefont{Sousa}},
  \bibinfo{author}{\bibfnamefont{P.}~\bibnamefont{Diaz}},
  \bibinfo{author}{\bibfnamefont{M.}~\bibnamefont{MacKenzie}},
  \bibinfo{author}{\bibfnamefont{J.}~\bibnamefont{Chapman}}, \bibnamefont{and}
  \bibinfo{author}{\bibfnamefont{P.}~\bibnamefont{Freitas}},
  \bibinfo{journal}{Eur. Phys. J. B} \textbf{\bibinfo{volume}{45}},
  \bibinfo{pages}{207} (\bibinfo{year}{2005}).

\bibitem[{\citenamefont{Teixeira et~al.}(2008)\citenamefont{Teixeira, Pereira,
  Ventura, Silva, Carpinteiro, Araujo, Sousa, Rickart, Cardoso, Ferreira
  et~al.}}]{Teixeira:EJPB:2008}
\bibinfo{author}{\bibfnamefont{J.~M.} \bibnamefont{Teixeira}},
  \bibinfo{author}{\bibfnamefont{A.~M.} \bibnamefont{Pereira}},
  \bibinfo{author}{\bibfnamefont{J.}~\bibnamefont{Ventura}},
  \bibinfo{author}{\bibfnamefont{R.~F.~A.} \bibnamefont{Silva}},
  \bibinfo{author}{\bibfnamefont{F.}~\bibnamefont{Carpinteiro}},
  \bibinfo{author}{\bibfnamefont{J.~P.} \bibnamefont{Araujo}},
  \bibinfo{author}{\bibfnamefont{J.~B.} \bibnamefont{Sousa}},
  \bibinfo{author}{\bibfnamefont{M.}~\bibnamefont{Rickart}},
  \bibinfo{author}{\bibfnamefont{S.}~\bibnamefont{Cardoso}},
  \bibinfo{author}{\bibfnamefont{R.}~\bibnamefont{Ferreira}},
  \bibnamefont{et~al.}, \bibinfo{journal}{Eur. Phys. J. B}
  \textbf{\bibinfo{volume}{82}}, \bibinfo{pages}{1486} (\bibinfo{year}{2008}).

\bibitem[{\citenamefont{Parkin et~al.}(1990)\citenamefont{Parkin, More, and
  Roche}}]{Parkin:PRL:1990}
\bibinfo{author}{\bibfnamefont{S.~S.~P.} \bibnamefont{Parkin}},
  \bibinfo{author}{\bibfnamefont{N.}~\bibnamefont{More}}, \bibnamefont{and}
  \bibinfo{author}{\bibfnamefont{K.~P.} \bibnamefont{Roche}},
  \bibinfo{journal}{Phys. Rev. Lett.} \textbf{\bibinfo{volume}{64}},
  \bibinfo{pages}{2304} (\bibinfo{year}{1990}).

\bibitem[{\citenamefont{Kiselev et~al.}(2004)\citenamefont{Kiselev, Sankey,
  Krivorotov, Emley, Rinkoski, Perez, Buhrman, and Ralph}}]{Kiselev:PRL:2004}
\bibinfo{author}{\bibfnamefont{S.~I.} \bibnamefont{Kiselev}},
  \bibinfo{author}{\bibfnamefont{J.~C.} \bibnamefont{Sankey}},
  \bibinfo{author}{\bibfnamefont{I.~N.} \bibnamefont{Krivorotov}},
  \bibinfo{author}{\bibfnamefont{N.~C.} \bibnamefont{Emley}},
  \bibinfo{author}{\bibfnamefont{M.}~\bibnamefont{Rinkoski}},
  \bibinfo{author}{\bibfnamefont{C.}~\bibnamefont{Perez}},
  \bibinfo{author}{\bibfnamefont{R.~A.} \bibnamefont{Buhrman}},
  \bibnamefont{and} \bibinfo{author}{\bibfnamefont{D.~C.} \bibnamefont{Ralph}},
  \bibinfo{journal}{Phys. Rev. Lett.} \textbf{\bibinfo{volume}{93}},
  \bibinfo{eid}{036601} (pages~\bibinfo{numpages}{4}) (\bibinfo{year}{2004}).

\bibitem[{\citenamefont{Braganca et~al.}(2005)\citenamefont{Braganca,
  Krivorotov, Ozatay, Garcia, Emley, Sankey, Ralph, and
  Buhrman}}]{Braganca:APL:2005}
\bibinfo{author}{\bibfnamefont{P.~M.} \bibnamefont{Braganca}},
  \bibinfo{author}{\bibfnamefont{I.~N.} \bibnamefont{Krivorotov}},
  \bibinfo{author}{\bibfnamefont{O.}~\bibnamefont{Ozatay}},
  \bibinfo{author}{\bibfnamefont{A.~G.~F.} \bibnamefont{Garcia}},
  \bibinfo{author}{\bibfnamefont{N.~C.} \bibnamefont{Emley}},
  \bibinfo{author}{\bibfnamefont{J.~C.} \bibnamefont{Sankey}},
  \bibinfo{author}{\bibfnamefont{D.~C.} \bibnamefont{Ralph}}, \bibnamefont{and}
  \bibinfo{author}{\bibfnamefont{R.~A.} \bibnamefont{Buhrman}},
  \bibinfo{journal}{Appl. Phys. Lett.} \textbf{\bibinfo{volume}{87}},
  \bibinfo{eid}{112507} (pages~\bibinfo{numpages}{3}) (\bibinfo{year}{2005}).

\bibitem[{\citenamefont{Petit et~al.}(2007)\citenamefont{Petit, Baraduc,
  Thirion, Ebels, Liu, Li, Wang, and Dieny}}]{Petit:PRL:2007}
\bibinfo{author}{\bibfnamefont{S.}~\bibnamefont{Petit}},
  \bibinfo{author}{\bibfnamefont{C.}~\bibnamefont{Baraduc}},
  \bibinfo{author}{\bibfnamefont{C.}~\bibnamefont{Thirion}},
  \bibinfo{author}{\bibfnamefont{U.}~\bibnamefont{Ebels}},
  \bibinfo{author}{\bibfnamefont{Y.}~\bibnamefont{Liu}},
  \bibinfo{author}{\bibfnamefont{M.}~\bibnamefont{Li}},
  \bibinfo{author}{\bibfnamefont{P.}~\bibnamefont{Wang}}, \bibnamefont{and}
  \bibinfo{author}{\bibfnamefont{B.}~\bibnamefont{Dieny}},
  \bibinfo{journal}{Phys. Rev. Lett.} \textbf{\bibinfo{volume}{98}},
  \bibinfo{eid}{077203} (pages~\bibinfo{numpages}{4}) (\bibinfo{year}{2007}).

\bibitem[{\citenamefont{Krivorotov et~al.}(2005)\citenamefont{Krivorotov,
  Emley, Sankey, Kiselev, Ralph, and Buhrman}}]{Krivorotov:Science:2005}
\bibinfo{author}{\bibfnamefont{I.~N.} \bibnamefont{Krivorotov}},
  \bibinfo{author}{\bibfnamefont{N.~C.} \bibnamefont{Emley}},
  \bibinfo{author}{\bibfnamefont{J.~C.} \bibnamefont{Sankey}},
  \bibinfo{author}{\bibfnamefont{S.~I.} \bibnamefont{Kiselev}},
  \bibinfo{author}{\bibfnamefont{D.~C.} \bibnamefont{Ralph}}, \bibnamefont{and}
  \bibinfo{author}{\bibfnamefont{R.~A.} \bibnamefont{Buhrman}},
  \bibinfo{journal}{Science} \textbf{\bibinfo{volume}{307}},
  \bibinfo{pages}{228} (\bibinfo{year}{2005}).

\bibitem[{\citenamefont{Lee and Lim}(2008)}]{Lee:APL:2008}
\bibinfo{author}{\bibfnamefont{D.~H.} \bibnamefont{Lee}} \bibnamefont{and}
  \bibinfo{author}{\bibfnamefont{S.~H.} \bibnamefont{Lim}},
  \bibinfo{journal}{Appl. Phys. Lett.} \textbf{\bibinfo{volume}{92}},
  \bibinfo{eid}{233502} (pages~\bibinfo{numpages}{3}) (\bibinfo{year}{2008}).

\bibitem[{\citenamefont{Devolder et~al.}(2009)\citenamefont{Devolder, Kim,
  Chappert, Hayakawa, Ito, Takahashi, Ikeda, and Ohno}}]{Devolder:JAP:2009}
\bibinfo{author}{\bibfnamefont{T.}~\bibnamefont{Devolder}},
  \bibinfo{author}{\bibfnamefont{J.-V.} \bibnamefont{Kim}},
  \bibinfo{author}{\bibfnamefont{C.}~\bibnamefont{Chappert}},
  \bibinfo{author}{\bibfnamefont{J.}~\bibnamefont{Hayakawa}},
  \bibinfo{author}{\bibfnamefont{K.}~\bibnamefont{Ito}},
  \bibinfo{author}{\bibfnamefont{H.}~\bibnamefont{Takahashi}},
  \bibinfo{author}{\bibfnamefont{S.}~\bibnamefont{Ikeda}}, \bibnamefont{and}
  \bibinfo{author}{\bibfnamefont{H.}~\bibnamefont{Ohno}},
  \bibinfo{journal}{J. Appl. Phys.} \textbf{\bibinfo{volume}{105}},
  \bibinfo{eid}{113924} (pages~\bibinfo{numpages}{5}) (\bibinfo{year}{2009}).

\bibitem[{\citenamefont{Mistral et~al.}(2006)\citenamefont{Mistral, Kim,
  Devolder, Crozat, Chappert, Katine, Carey, and Ito}}]{Mistral:APL:2006}
\bibinfo{author}{\bibfnamefont{Q.}~\bibnamefont{Mistral}},
  \bibinfo{author}{\bibfnamefont{J.-V.} \bibnamefont{Kim}},
  \bibinfo{author}{\bibfnamefont{T.}~\bibnamefont{Devolder}},
  \bibinfo{author}{\bibfnamefont{P.}~\bibnamefont{Crozat}},
  \bibinfo{author}{\bibfnamefont{C.}~\bibnamefont{Chappert}},
  \bibinfo{author}{\bibfnamefont{J.~A.} \bibnamefont{Katine}},
  \bibinfo{author}{\bibfnamefont{M.~J.} \bibnamefont{Carey}}, \bibnamefont{and}
  \bibinfo{author}{\bibfnamefont{K.}~\bibnamefont{Ito}},
  \bibinfo{journal}{Appl. Phys. Lett.} \textbf{\bibinfo{volume}{88}},
  \bibinfo{eid}{192507} (pages~\bibinfo{numpages}{3}) (\bibinfo{year}{2006}).

\bibitem[{\citenamefont{Worledge}(2004{\natexlab{a}})}]{Worledge:APL:2004b}
\bibinfo{author}{\bibfnamefont{D.~C.} \bibnamefont{Worledge}},
  \bibinfo{journal}{Appl. Phys. Lett.} \textbf{\bibinfo{volume}{84}},
  \bibinfo{pages}{4559} (\bibinfo{year}{2004}{\natexlab{a}}).

\bibitem[{\citenamefont{Worledge et~al.}(2007)\citenamefont{Worledge,
  Trouilloud, and Gallagher}}]{Worledge:APL:2007}
\bibinfo{author}{\bibfnamefont{D.~C.} \bibnamefont{Worledge}},
  \bibinfo{author}{\bibfnamefont{P.~L.} \bibnamefont{Trouilloud}},
  \bibnamefont{and} \bibinfo{author}{\bibfnamefont{W.~J.}
  \bibnamefont{Gallagher}}, \bibinfo{journal}{Appl. Phys. Lett.}
  \textbf{\bibinfo{volume}{90}}, \bibinfo{eid}{222506}
  (pages~\bibinfo{numpages}{3}) (\bibinfo{year}{2007}).

\bibitem[{\citenamefont{Worledge}(2004{\natexlab{b}})}]{Worledge:APL:2004}
\bibinfo{author}{\bibfnamefont{D.~C.} \bibnamefont{Worledge}},
  \bibinfo{journal}{Appl. Phys. Lett.} \textbf{\bibinfo{volume}{84}},
  \bibinfo{pages}{2847} (\bibinfo{year}{2004}{\natexlab{b}}).

\bibitem[{\citenamefont{Devolder
  et~al.}(2008{\natexlab{b}})\citenamefont{Devolder, Hayakawa, Ito, Takahashi,
  Ikeda, Crozat, Zerounian, Kim, Chappert, and Ohno}}]{Devolder:PRL:2008}
\bibinfo{author}{\bibfnamefont{T.}~\bibnamefont{Devolder}},
  \bibinfo{author}{\bibfnamefont{J.}~\bibnamefont{Hayakawa}},
  \bibinfo{author}{\bibfnamefont{K.}~\bibnamefont{Ito}},
  \bibinfo{author}{\bibfnamefont{H.}~\bibnamefont{Takahashi}},
  \bibinfo{author}{\bibfnamefont{S.}~\bibnamefont{Ikeda}},
  \bibinfo{author}{\bibfnamefont{P.}~\bibnamefont{Crozat}},
  \bibinfo{author}{\bibfnamefont{N.}~\bibnamefont{Zerounian}},
  \bibinfo{author}{\bibfnamefont{J.-V.} \bibnamefont{Kim}},
  \bibinfo{author}{\bibfnamefont{C.}~\bibnamefont{Chappert}}, \bibnamefont{and}
  \bibinfo{author}{\bibfnamefont{H.}~\bibnamefont{Ohno}},
  \bibinfo{journal}{Phys. Rev. Lett.} \textbf{\bibinfo{volume}{100}},
  \bibinfo{eid}{057206} (\bibinfo{year}{2008}{\natexlab{b}}).
\end{thebibliography}

\begin{appendix}

\section{Derivation of magneto-resistance noise signature}
\label{sec:appendix-sensitivity}

%%%%%%%%%%%%%%%%%%%%%%%%%%%%%%%%%%%%%%%%%%%%%%%%%%%%%%%%%%%%%%%%%%%
In this appendix we derive eq.~\eqref{eq:MRnoise-princ} and the leading order terms in Table~\ref{tab:sensitivity} starting from eq.~\eqref{eq:deltaR}.\\

The first step is to evaluate the dot product in \eqref{eq:deltaR} of the dynamical magnetization $\delta {\vec M}_F^U(\vec k)$ of the FL with the equilibrium magnetization ${\vec M}_1(\vec r)$ of the RL. Since $\delta {\vec M}_F^U(\vec k)$ is perpendicular to the macrospin component ${\vec M}_F^U$, the dot product of $\delta {\vec M}_F^U(\vec k)$ with ${\vec M}_1^U$ can be expressed in terms of the angle $\theta$ between the macrospins ${\vec M}_{F}^U$ and ${\vec M}_{1}^U$ of the two layers. Similarly, decomposing $\delta {\vec M}_1^E(\vec r)$ into a component $\delta {M}_{1,\parallel}^E(\vec r)$ parallel to the macrospin ${\vec M}_1^U$, and a component $\delta {M}_{1,\perp}^E(\vec r)$ perpendicular to ${\vec M}_1^U$, allows to evaluate the dot product of $\delta {\vec M}_1^E(\vec r)$ with $\delta {\vec M}_F^U(\vec k)$. With that the MR variation~\eqref{eq:deltaR} becomes
\begin{equation}
\delta R_F(\vec k,\theta) = \delta {M}_F^U(\vec k)\,\left[\overline {M_1^V}(\vec k,\theta)+\,\overline {M_1^{E}}(\vec k,\theta)\right]\label{eq:MRnoise}
\end{equation}
where
\begin{equation}
\overline {M_1^V}(\vec k,\theta) = \sin \theta\,{M_1}\, {\overline{W}_x}(n_{x},\phi_{x})\,{\overline{W}_y}(n_{y},\phi_{y}), \label{eq:modechar}
\end{equation}
%%%%%%%
\begin{equation}
{\overline{W}_x}(n_{x},\phi_{x})\,{\overline{W}_y}(n_{y},\phi_{y})=\frac{1}{S_{pil}} \int_{S_{pil}}\Re{e^{i \vec k \vec r}} \, d \vec r, \label{eq:averageWF}
\end{equation}
and
\begin{eqnarray}
\lefteqn{\overline {M_1^{E}}(\vec k,\theta)=}\nonumber\\
&&\frac{1}{S_{pil}} \int_{S_{pil}}\left[\cos \theta\,\delta {M}_{1,\perp}^E(\vec r)+\sin \theta\,\delta {M}_{1,\parallel}^E(\vec r)\right]  \Re{e^{i \vec k \vec r}}\,d \vec r.\nonumber\\ \label{eq:magconf}
\end{eqnarray}
As the edge domain contributions~\eqref{eq:magconf} are relevant only on the EA in P and AP state where $\theta$ is basically $0$ or $\pi$, the term with $\sin \theta$ in~\eqref{eq:magconf} is in all practical cases negligible, such that only the term $\delta {M}_{1,\perp}^E(\vec r)=\delta {M}_{1,y}^E(x,y)$ remains. \\
In the following, we derive the leading order terms of these quantities listed in Table~\ref{tab:sensitivity}.

\subsection{Micromagnetic configuration dependence}
\label{sec:appendixA-micro}

Decomposing $\theta$ into the angle $\theta_0$ between the two macrospins for an ideal pillar, and a small deviation $\Delta\theta$ due to misalignments, $\sin \theta$ and $\cos \theta$ in $\overline {M_1^V}(\vec k,\theta)$ and $\overline {M_1^{E}}(\vec k,\theta)$ can be expanded in $\Delta\theta$ about $\theta_0$, where the leading order terms for the different field regions are summarized in Tab.~\ref{tab:sensitivity}(a). The underlying values of $\theta_0$ are as follows:
For EA applied field, $\theta_0=0$ in the P state, $\theta_0=\pi$ in the AP state, and $0\ll\theta_0<\pi$ above the SF at positive fields, and below the 2nd SF at negative fields. For HA field, $\theta_0$ decreases continuously from $\pi$ at zero field to a value close to $\pi/2$ at the saturation field of the FL, and finally towards zero as the RL magnetization continues to tilt towards the HA. \\

The presence of edge domains on the EA can be deduced from the hysteresis loops: deviations of the resistance from its saturation values in P and AP state indicate non-uniformities of the FL and/or the RL magnetization. E.g. in the P state, the resistance
increases continuously when the (ascending) field approaches the switching field to the AP state, both magnetizations being subject to an increasing effective antiparallel field consisting of the (self)demagnetizing field, the mutual dipolar coupling field, and the external field as soon as it becomes positive. \\
In the AP state, at low negative field just before the switching to the P state, both the external field and the interlayer dipolar coupling field are parallel to the magnetization of the RL, thus partly suppressing the edge domains created by the (self)demagnetizing field; in contrast, edge domains in the FL are only suppressed by the interlayer dipolar coupling field, but enhanced by the external field. Indeed, the resistance in the low-field AP sate departs much less from AP remanence than it does from the P remanence in the low-field P state. \\
Finally, in the high-field AP state, the resistance starts to decrease continuously already long before the SF transition due to increasing non-uniformities of the RL magnetization, which is pointing antiparallel to the high external field.\\ On the HA, edge domain contributions are negligible compared to the zeroth order contributions of the volume magnetization, except for zero field where the macrospins are antiparallel.
 
\subsection{Mode character dependence}
\label{sec:appendixA-mode}

By means of eqs. \eqref{eq:sines} and \eqref{eq:defn} the two integrals  ${\overline{W}_x}$ and ${\overline{W}_y}$ over the layer dimension in direction $x$ and $y$, respectively, are easily evaluated as
\begin{equation}
{\overline{W}_x}(n_{x},\phi_{x})= 
\frac{2}{n_{x}\pi}\sin(n_{x}\frac{\pi}{2})\sin\phi_{x},\label{eq:int-wfx}
\end{equation}
where ${\overline{W}_y}(n_{y},\phi_{y})$ is given by an analogous expression. \\
Decomposing the mode numbers and the phase as in section~\ref{sec:model3L}, $n_{x}=n_{x}^0+\Delta n_{x}$ and $\phi_{x}=\phi_{x}^0+\Delta\phi_{x}$ where $\phi_{x}^0=\pi/2+n_{x}^0\cdot\pi/2$ is the phase for symmetric pinning, ${\overline{W}_x}(n_{x},\phi_{x})$ can be expanded in   $\Delta\phi_{x}\ll1$ and either $\Delta n_{x}\ll1$ (weak pinning) or $(1-\Delta n_{x})\ll1$ (strong pinning). The result as a function of $n_{x}$ is shown in Table \ref{tab:sensitivity}(b).\\

The edge domain contributions $\overline {\delta {M}_l^E}$ to the MR noise in the P and AP state can be evaluated on the basis of symmetry considerations. In spite of a non-uniform equilibrium magnetization, $\overline {\delta {M}_l^E}$ is zero if the  product of the wavefunction and the function describing the spatial dependence of the $y$-component of the edge domain magnetization under the integral~\eqref{eq:magconf} is either zero or antisymmetric in $x$- or $y$-coordinate. For strong pinning, the product of the two functions is zero (or negligibly small), because near the layer edges, where the edge domain magnetization is non-zero, the wavefunction has minimum amplitude due to the pinning. Significant contributions from edge domains can be expected for weak pinning only. In this case, the integral~\eqref{eq:magconf} will vanish for certain modes if the magnetization for a given micromagnetic state is invariant under reflection or rotation or a combination of both. The flower-state e.g. is invariant under reflections about $x$- and $y$-axis, i.e. the $y$-component of the magnetization is antisymmetric in both $x$- and $y$-coordinate. $\overline {\delta {M}_l^E}$ is therefore non-zero only for modes with two odd mode numbers. Similarly, it can be shown that for the S-state, $\overline {\delta {M}_l^E}$ is non-zero for modes, whose mode numbers are either both odd or both even; the C-state renders modes with odd $n_{x}$ visible.\\

%(even, even) & (0,0), (2,0)  & S-state\\
%(even, odd)   & (0,1), (2,1)  & -- \\
%(odd, even)   & (1,0), (3,0)  & C-state \\
%(odd, odd)   & (1,1)  & S-, C-, and flower-state 

Finally, we derive the expected relative intensity of the modes. $\Delta n_{x}$ for weak pinning or $(1-\Delta n_{x})$ for strong pinning are of the order 0.1. We may assume that for small asymmetries of the pinning, $\Delta\phi_{x}$ is at most of the same order of magnitude as $\Delta n_{x}$ (or $(1-\Delta n_{x})$).
Therefore, for weak pinning,  ${\overline{W}_x}(n_{x}^0\geq1)\approx0.1/n_{x}$ is one order of magnitude smaller than ${\overline{W}_x}(n_{x}^0=0)\approx1$, and we expect to observe in addition to the quasi-uniform mode close to (0,0) higher modes with mode numbers close to (1,0), (0,1), (2,0), (0,2), and possibly (3,0). Their intensities, being proportional to $({\overline{W}_x}{\overline{W}_y})^2$, scale with factors quadratical in $\Delta\phi_{x,y}$ or $\Delta n_{x,y}$, and are therefore two orders of magnitude lower than that of (0,0). The intensities of all other modes, such as (1,1), are of forth order in $\Delta\phi_{x,y}$ and $\Delta n_{x,y}$, or strongly reduced by the factor $1/(n_x n_y)^2$, and therefore most likely too weak to be detected.\\
For strong pinning and even $n_{x}^0$, ${\overline{W}_x}(n_{x}^0\geq1)\approx1/n_{x}$ is of the same order of magnitude as ${\overline{W}_x}(n_{x}^0=0)\approx1$. In contrast, for odd $n_{x}^0$, ${\overline{W}_x}(n_{x}^0\geq1)\approx0.01/n_{x}$ is two orders of magnitude smaller than ${\overline{W}_x}(n_{x}^0=0)$. Consequently, the higher modes close to (3,1), (1,3), and (5,1) will have intensities comparable to that of the fundamental mode close to (1,1), or one order of magnitude lower due to the factor $1/(n_x n_y)^2$.

\section{Details on extraction of model parameters}
\label{sec:appendix-extrapara}

%%%%%%%%%%%%%%%%%%%%%%%%%%%%%%%%%%%%%%%%%%%%%%%%%%%%%%%%%%%%%%%%%%%%%%%%%%%%%%%%%%%%%%

In this annex, we present the arguments used to extract the  
material parameters $M_l$, $A_l$, $J^{eb}$, $J^{int}$, the geometry parameters $(N_l^{x}$, $N_l^{y}$, $N_l^{z})$, and $(N_{ml}^{x}$, $N_{ml}^{y}$, $N_{ml}^{z})$, as well as the mode numbers $(n_x,n_y)$
from the experimental spectra. 

\subsection{Reduction of number of free parameters}

Given the (approximate) layer dimensions $L_x, L_y, L_z$, the demagnetizing factors $N_l^{x}$, $N_l^{y}$, $N_l^{z}$ can be calculated using e.g. OOMMF simulations, where we find $N_l^{z}\approx 1-(N_{l}^{x}+N_{l}^{y})$ and $N_l^{y}/N_{l}^{x}\approx L_x/L_y$ as should be expected. Using the formulae in Ref.~\onlinecite{Newell:JGP:1993} it can be shown that the dipolar coupling constants obey similar relations, $N_{ml}^{y}/N_{ml}^{x}=L_x/L_y$ and $N_{ml}^{z}=-(N_{ml}^{x}+N_{ml}^{y})$, and for symmetry reasons, $\mathbf{N}_{ml}=\mathbf{N}_{lm}$. The remaining components $N_{F1}^{x}$, $N_{F2}^{x}$, and $N_{12}^{x}$ are kept as free parameters to be extracted from the experiment, although they can be calculated by means of Ref.~\onlinecite{Newell:JGP:1993}. \\
On the basis of previous measurements on MTJ stacks, the number of free parameters can be further reduced: 
In Ref.~\onlinecite{Kubota:APL:2006} it has been shown that the magnetization of the CoFeB free layer does not depend on the layer thickness in the range from 2 to 3 nm. We may therefore assume that the FL and the RL - being of the same material, but having different thicknesses - have equal magnetizations, $M_1=M_F$. Moreover, we expect the layer magnetizations in the pillar to be reduced for all layers by the same (relative) amount w.r.t. the thin film saturation magnetizations: $M_2^{pillar}/M_2^{film}=M_{F,1}^{pillar}/M_{F,1}^{film}$. 
Analogous relations are expected to hold for the exchange stiffness constants $A_l$: $A_1=A_F$ and $A_2^{pillar}/A_2^{film}=A_{F,1}^{pillar}/A_{F,1}^{film}$. 
%Altogether, we are left with the following free parameters: $M_F$, $A_F$, $J^{int}$, $J^{eb}$, $N_{F1}^{x}$, $N_{F2}^{x}$, $N_{12}^{x}$, and the mode numbers of the lowest mode. The mode numbers of the higher modes are fixed as soon as those of the lowest mode are (see section~\ref{sec:model3L}).

%%%%%%%%%%%%%%%%%%%%%%%%%%%%%%%%%%%%%%%%%%%%%%%%%%%%%%%%%%%%%%%%%%%%%%%%%%%%%%%%%%%%

\subsection{Literature values}
\label{sec:appendix-literature}

In this paragraph, we list as an orientation literature values for the material parameters.\\
As thin film exchange stiffness constants we use the values of the 40~nm CoFeB and CoFe films in Ref. ~\onlinecite{Bilzer:JAP:2006}:  $A_{F,1}^{film}=28.4\times 10^{-12}\textrm{J}/\textrm{m}$ and $A_2^{film}=27.5 \times 10^{-12}\textrm{J}/\textrm{m}$.\\
The magnetizations of CoFe and annealed CoFeB depend on the percentage of Fe in Co: The bulk value for both $\textrm{Co}_{70}\textrm{Fe}_{30}$ and $\textrm{Co}_{75}\textrm{Fe}_{25}$ (corresponding to $\textrm{Co}_{60}\textrm{Fe}_{20}\textrm{B}_{20}$) is $(2.2\pm0.1)$~T,\cite{Bozorth:book:1993,Lamy:JAP:2005}. The (thin film) free layer magnetization of our MTJ stack has been measured to be $\mu_0 M_{F}^{film}=(1.8\pm0.1)$~T.\cite{Cornelissen:JAP:2009} The thin film value for the CoFe layer is expected to be in the interval $\mu_0 M_2^{film}=(2.0\pm0.2)$~T.\\
The exchange bias field in a $\textrm{Co}_{90}\textrm{Fe}_{10}$~(5~nm)/ PtMn~(20~nm) system has been measured to be $\mu_0 H^{eb}\approx 67$~mT,\cite{Rickart:EJPB:2005,Teixeira:EJPB:2008} which corresponds to an exchange bias energy  of $J^{eb}= 4.5 \times 10^{-4}\textrm{J}/\textrm{m}^2$, using 2.0~T as saturation magnetization of the CoFe layer. For the interlayer exchange energy a maximum value of $J^{int}=-5\times 10^{-4}\textrm{J}/\textrm{m}^2$ has been reported.\cite{Parkin:PRL:1990}\\
Ref.~\onlinecite{Newell:JGP:1993} allows to calculate the dipolar coupling constant for two rectangular layers of equal thicknesses. As in our pillars the FL has a different thickness than the two SAF layers, only $N_{12}^{x}$ may be calculated directly, yielding $N_{12}^{x}=0.016$. The dipolar coupling constants $N_{F1}^{x}$ and $N_{F2}^{x}$ involving the FL can only be estimated as the mean value of the constants calculated for two 3~nm thick layers and for two 2~nm thick layers, from which we obtain $N_{F1}^{x}\approx 0.018$ and $N_{F2}^{x}\approx 0.013$ (maximum deviation $\pm0.003$). 

\subsection{Regression method}

$M_F$ is determined by the modes F0 on the EA with a weak dependence on the chosen mode numbers of F0 (see discussion below). A minimum value for $M_F$ of 1.25~T follows from the measured room-temperature anisotropy field, which must be smaller than the calculated (zero-temperature) anisotropy field. $M_1$ and $M_2$ cannot be extracted directly, but depend entirely on the above assumptions. Once $M_1$ and $M_2$ have been fixed, $J^{int}$ and $J^{eb}$ can be estimated from the spin-flop field and the mode A0 on the HA. $N_{F1}^{x}$ and $N_{F2}^{x}$ follow from the gap opening in the mode F0 on the EA and the shift of the EA hysteresis loop to negative fields: We have used the above calcuted values for $N_{F1}^{x}$, $N_{F2}^{x}$, and $N_{12}^{x}$ as starting values, which we have adapted to the experimental data by rescaling, assuming that the deviation of the mutual dipolar fields, e.g. due to micromagnetics, is similar for all pillar layers. It turns out that $N_{F1}^{x}$ as the largest component can be maximum 0.01, because otherwise the gap opening exceeds the observed 2~GHz (maximum value of 2.5~GHz for the calculated coefficients). On the other hand, the difference of $N_{F1}^{x}$ and $N_{F2}^{x}$ must be at least $0.005$ to ensure a shift of the hysteresis loop of minimum $5$~mT. $N_{F1}^{x}=0.01$ and $N_{F2}^{x}=0.005$ are therefore uniquely determined (maximum deviation $\pm0.002$). Since for the calculated constants $N_{12}^{x}$ is between $N_{F1}^{x}$ and $N_{F2}^{x}$, we set $N_{12}^{x}=0.007$. \\ 

The exchange stiffness constant $A_F$ and the mode numbers $(n_x,n_y)$ cannot be extracted separately, since they enter the effective field (and consequently the frequencies) only as a product. The BC in the pillar being unknown, the lowest modes can have any mode numbers between (0,0) (unpinned BC) and (1,1) (totally pinned BC), where $n_x$ can be larger than $n_y$ (cf. section~\ref{sec:model3L}). \\
In order to adjust F0 on the EA with the mode (1,1) in the limit of strong pinning, we would need $A_F\approx 1/20\cdot A_{F}^{film}$ and $\mu_0  M_F\approx 1.1$~T; this value for $M_F$ is significantly smaller than the allowed minimum, and the reduction of $A_F$ w.r.t. its thin film value is unreasonably large given that $M_F\approx 2/3 \cdot M_{F}^{film}$. In addition, a discrepancy of more than 1.5~GHz between calculated and measured mode F0 is observed on the HA even at high fields. 
Similarly, fitting F0 with the mode (0.5,0.5), which might be considered as the border between strong and weak pinning for $n_x=n_y$, or the mode (1,0), for maximum difference $n_x-n_y$, still requires $A_F< 1/5\cdot A_{F}^{film}$ and $\mu_0 M_F\approx 1.2$~T. Consequently, strong pinning can be excluded in our pillars; the mode numbers of F0 must be well below (0.4,0.4) or (0.8,0). The pinning is weak. This is also corroborated by the fact that in particular on the HA the lowest mode F0 has much higher intensity than the higher modes F1 to F5, which is a characteristics of weakly pinned systems (cf. section~\ref{sec:sensitivity}).\\
If we assume totally unpinned BC - fitting F0 with (0,0) and the higher modes with (1,0), (0,1) etc. - we get $\mu_0  M_F=1.3$~T and $A_F\approx 2/3\cdot A_{F}^{film}$, i.e. approximately $A_F\varpropto M_F$. To narrow the mode numbers down within these borders, we assume that indeed $n_x>n_y$, which finally confines $(n_x,n_y)$ on the HA to $(n_x,n_y)\leq(0.2,0.1)$, and on the EA to $(n_x,n_y)\leq(0.4,0.2)$. To fit F0 on EA and HA simultaneously, requires that the mode numbers on the HA are smaller than on the EA, as had already been suggested in section~\ref{sec:model3L}.

\end{appendix}

\end{document}